\documentclass[12pt]{iopart}
\usepackage{graphicx}% Include figure files
\usepackage{dcolumn}%Align table columns on decimal point
\usepackage{bm}% bold math

\usepackage{amssymb}

\newcommand{\sm}{{\cal \bf S2M\,}}
\newcommand{\im}{{\cal \bf I2M\,}}
\newcommand{\npse}{{\cal \bf NPSE\,}}
\newcommand{\gpod}{{\cal \bf GP1D\,}}
\newcommand{\tdgpetd}{{\cal \bf GP3D\,}}
\newcommand{\tdgpeod}{{\cal \bf GP1D\,}}

\begin{document}
\title[Weakly linked binary mixtures of 
$F=1$ $^{87}$Rb Bose-Einstein condensates ]
{Weakly linked binary mixtures of 
$F=1$ $^{87}$Rb Bose-Einstein condensates}

\author{M. Mel\'e-Messeguer$^1$, 
B. Juli\'a-D\'\i az$^{1,2}$, 
M. Guilleumas$^1$, 
A. Polls$^{1,2}$, 
A. Sanpera$^{3,4}$} 

\address{$^1$ Departament d'Estructura 
i Constituents de la Mat\`{e}ria,\\ 
Universitat de Barcelona, E--08028, 
Barcelona, Spain.}

\address{$^2$ Institut de Ci\`{e}ncies del Cosmos,
Universitat de Barcelona, E--08028 Barcelona, Spain.}

\address{$^3$ ICREA-Instituci\'o 
Catalana de Recerca i Estudis 
Avan\c{c}ats, Llu\'{i}s 
Companys 23, E--08010 Barcelona, Spain.}

\address{$^4$
Grup de F\'{\i}sica Te\`orica: Informaci\'o i 
Processos Qu\`antics. Universitat Aut\`onoma 
de Barcelona, E--08193 Bellaterra, Spain.}

\ead{marina@ecm.ub.es}

\begin{abstract}

We present a study of binary mixtures of Bose-Einstein condensates confined in
a double-well potential within the framework of the mean field
Gross-Pitaevskii equation. We reexamine both the single component and
the binary mixture cases for such a potential, and we investigate in which
situations a simpler two-mode approach leads to an accurate description
of their dynamics. We also estimate the validity of the most usual
dimensionality reductions used to solve the Gross-Pitaevskii equations. To
this end, we compare both the semi-analytical two-mode approaches and the
numerical simulations of the 1D reductions with the full 3D numerical
solutions of the Gross-Pitaevskii equation. Our analysis provides a guide to
clarify the validity of several simplified models that describe mean field
non-linear dynamics, using an experimentally feasible binary mixture of an
$F=1$ spinor condensate with two of its Zeeman manifolds populated, $m=\pm 1$. 
\end{abstract}

\pacs{
03.75.Mn %multicomponent and spinor condensates
03.75.Lm %Tunneling, Josephson effect, Bose-Einstein condensates in periodic potentials, solitons, vortices, and topological excitations
03.75.Kk %Dynamic properties of condensates; collective and hydrodynamic excitations, superfluid flow
74.50.+r %Tunneling phenomena; point contacts, weak links, Josephson effects (for SQUIDs, see 85.25.Dq; for Josephson devices, see 85.25.Cp; for Josephson junction arrays, see 74.81.Fa)}
}
\maketitle

\tableofcontents

\section{Introduction}

The phase coherence of a Bose-Einstein Condensate (BEC) 
is an important and characteristic property of ultracold 
bosonic gases that leads to fascinating macroscopic 
phenomena such as interference effects or Josephson-type 
oscillations. Two condensates trapped in a double-well 
potential exhibit interference fringes when the barrier 
is released and the two expanding condensates, with 
a well-defined quantum phase, overlap. Instead, if 
the barrier is not switched-off and is large enough 
to ensure a weak link between both condensates in 
each side of the trap, the quantum phase difference 
will drive Josephson-like effects, which consist on 
fast oscillating tunneling, much faster than the 
single particle tunneling, of atoms through the 
potential barrier~\cite{Leggett01,pitaevski2003}.

The first evidence of the phase coherence of a BEC 
was obtained in early interference 
experiments~\cite{interference} where clean 
interference patterns appeared in the overlapping 
region of two expanding condensates. It has been 
only recently that a clear evidence of a bosonic 
Josephson junction in a weakly linked 
scalar BEC~\footnote{The notation ``scalar BEC'' 
is used as equivalent to ``single component BEC'' 
in this article.} has been experimentally reported by the 
group of M. Oberthaler in Heidelberg~\cite{Albiez05}. 
In this experiment, two condensates are confined 
in a double-well potential with an initial population 
imbalance between both sides which 
triggers the Josephson oscillations. The 
tunneling of particles leads to a coupled 
dynamical evolution of the two conjugate 
variables, the phase difference between the 
two weakly linked condensates and their population 
imbalance. In spite of the system being very 
dilute, the inter-species interaction plays a 
crucial role in the Josephson dynamics, leading 
to new regimes beyond the standard Josephson 
effect, e.g. macroscopic quantum self trapping 
(MQST).

The Gross-Pitaevskii (GP) mean field theory 
provides a natural framework for investigating 
Josephson dynamics in weakly interacting systems 
at very low temperature. Josephson oscillations 
in scalar Bose-Einstein condensates have been 
theoretically studied by using different 
techniques~\cite{Smerzi97,Raghavan99,Milburn97, foerster05,
Ananikian2006,Gati2007,jame05,Cirac98,java99,tripp08,cederbaum09,
ours10,martorell2010,Carr2010,Vardi2009}. 
As expected, the full three dimensional time-dependent 
Gross-Pitaevskii equation (\tdgpetd) provides 
an excellent agreement with the 
experimental data~\cite{Ananikian2006,Gati2007,AlbiezPhD}. 
However, since 3D dynamics need in general 
rather involved calculations, one can benefit 
from the fact that the barrier is created 
along one direction and the tunneling of 
particles is mainly one dimensional (1D) to 
investigate the Josephson dynamics by means 
of effective 1D GP-like equations. Among 
these reduced GP equations, the non-polynomial 
nonlinear Schr\"odinger equation (\npse) 
proposed in Ref.~\cite{Salasnich2002} has 
provided the best agreement with the experimental 
results in scalar condensates~\cite{AlbiezPhD}, 
whereas another effective 1D Gross-Pitaevskii 
equation ( \gpod) fail to describe the dynamics for 
large number of trapped atoms in the same 
trapping conditions as in the Heidelberg's 
experiment~\cite{Gati2007,AlbiezPhD}.

Interactions are important to understand  the different 
regimes of the tunneling dynamics.
Therefore, multi-component BECs in 
double-well potentials offer an interesting 
extension to study phenomena related to 
phase coherence. In particular the Josephson 
dynamics will become richer due to the 
interplay between intra- and 
inter-species interactions.

Josephson oscillations in binary mixtures confined in 
double-well potentials have been addressed in 
a number of recent articles. The case of 
two-component BECs with density-density 
interactions has been studied within two-mode 
approaches in Refs.~\cite{ashab02,2component,xu2008,Satija2009,
Mazzarella2009,Sun2009,naddeo10,Mazzarella2010}. 
Refs.~\cite{Mazzarella2009,Mazzarella2010} go one 
step further and also consider \tdgpeod simulations. 
Spin-dependent interactions have been addressed in 
Refs.~\cite{Bruno2009, Wang2010}.
Josephson dynamics in 
spinor condensates confined in double-wells, characterized by an 
exchange of population between different 
Zeeman components, has also been investigated 
in Refs.~\cite{spin,Mele2009}. In Refs.~\cite{Mazzarella2009,Bruno2009} 
the interest of studying Josephson dynamics 
in binary mixtures has been emphasized as it can 
give access to information of the different 
scattering lengths present in the system.

Recently, the equations of the tunneling dynamics in a 
binary mixture within the 
two-mode approximation to the GP equations 
have been derived in Ref.~\cite{Satija2009}. 
However, the authors have not compared their 
two-mode analysis to direct numerical 
resolutions of the GP equation, and have also 
not provided microscopic values to 
the parameters of the two-mode equations. 
Their main result is the description of a
symmetry breaking pattern occurring when the 
inter- and intra-species interactions differ 
substantially. In Ref.~\cite{Mazzarella2009} 
a comparison of the standard two-mode approach 
and the coupled \gpod equations for the mixture 
has been presented for one specific double-well potential which allows an analytical 
treatment. 

For single component BECs, it has been already studied 
the range of validity of the different approximations to the Josephson
dynamics by comparing with \tdgpetd calculations and 
with the experimental results. However,
no comparison with the full \tdgpetd dynamics has been yet performed
for a binary mixture in a double-well potential.

The aim of this paper is to investigate systematically 
the tunneling dynamics of a 
binary mixture of BECs trapped in 
a double-well potential, as well as the validity of the different mean-field 
approximations.
We consider a mixture of two components obtained 
by populating two Zeeman states 
of an $F=1$ $^{87}$Rb condensate confined in the
same double-well potential as in the experiments~\cite{Albiez05}.
This system corresponds to a natural extension of 
the experimental work of Ref.~\cite{Albiez05}, where 
only one of the Zeeman components was populated. 

We provide a general overview of
the different techniques used to investigate Josephson dynamics 
within the two-mode model (standard and improved two-mode)  and 
within the Gross-Pitaevskii framework (one-dimensional
reductions of the GP equation, \gpod and \npse).
To this end,  we solve the full 3D time-dependent GP equation for the mixture
as a reference  to assess and analyze the validity of the previous approximations.

The paper is organized as follows. The general 
framework of the coupled Gross-Pitaevskii equations 
for a binary mixture is presented in 
Sec.~\ref{sec:system}. In Sec.~\ref{sec:tw}, we 
derive analytic two-mode models both for single and 
two component systems. First we  recall the 
standard two-mode model (\sm). Then we derive 
the equations of the improved two-mode model (\im)
for a binary mixture, generalizing the work for 
a single component BEC performed in Ref.~\cite{Ananikian2006}.
We discuss the stability of the dynamical equations and 
look for the stationary points for a binary mixture.
In Sec.~\ref{3D-1D-sec}, we analyze the different one-dimensional
reductions of the \tdgpetd equations for the mixture: \gpod and \npse.  
In Sec.~\ref{sec:sc}, we revisit the dynamics of a single component
condensate in a double-well potential with the same parameters as
in the experiment~\cite{Albiez05}. 
The tunneling dynamics in two-component systems is 
accurately discussed in Sec.~\ref{sec:bmsp}. 
We obtain the dynamics by solving the coupled \tdgpetd equations 
for the mixture and show that for certain conditions there exists a good agreement between
\im and  \tdgpetd, as well as for \npse and \tdgpetd.
The range of validity of the two-mode models is explored, paying special
attention to situations which fall beyond the two-mode approximation.
Finally, we discuss cases that present characteristic 
features arising from the mixture, with no analog in the 
tunneling dynamics of a single component BEC.
Conclusions are given in Sec.~\ref{sec:sum}.

\section{Mean field approach: Gross-Pitaevskii equations}
\label{sec:system}

We consider a binary mixture of weakly interacting 
%Bose-Einstein condensates
atoms at zero temperature, 
confined by the same double-well potential, $V({\bf r})$. 
For dilute systems with sufficiently large number of 
particles, the Gross-Pitaevskii equation 
provides a suitable framework to study the dynamics. 
In the mean field approximation, each condensate 
is described by the corresponding wave function 
$\Psi_i({\bf r};t)$, with $i=a,b$ denoting each 
of the two components of the binary mixture. To avoid 
any misunderstanding let us remind the reader that 
we are describing two different kind of atoms, 
$a$ and $b$, which evolve on a double-well 
external potential. In most situations, the 
system will behave as if there were four weakly 
linked Bose-Einstein condensates, two per each 
component of the binary mixture per each 
side of the potential barrier. The mean field 
description will reflect this feature by the 
homogeneous quantum phase of $\Psi_i({\bf r};t)$ 
at each side of the potential barrier, as will 
be discussed in great detail in the following 
sections. 

The dynamical evolution of the two wave functions 
can be obtained by solving the two coupled GP 
equations:

\begin{center}
\begin{eqnarray}
i\hbar{\partial \Psi_a({\bf r};t)\over \partial t} &=&
\left[-{\hbar^2\over 2 m_a}\nabla^2
+V({\bf r}) +g_{aa} N_a |\Psi_a({\bf r};t)|^2 
+g_{ab} N_b |\Psi_b({\bf r};t)|^2 \right]\Psi_a({\bf r};t)\nonumber \\
i\hbar{\partial \Psi_b({\bf r};t)\over \partial t} &=&
\left[-{\hbar^2\over 2 m_b}\nabla^2
+V({\bf r}) +g_{ba} N_a |\Psi_a({\bf r};t)|^2 
+g_{bb} N_b|\Psi_b({\bf r};t)|^2\right] \Psi_b ({\bf r};t)\,.\nonumber \\
\label{GP-mixture}
\end{eqnarray}
\end{center}
For each component, the condensate wave function 
$\Psi_i({\bf r};t)$ is normalized to 1, $m_i$ is 
the atomic mass, and $g_{ii}=4 \pi \hbar^2 a_i/m_i$ is 
the effective atomic interaction between atoms of the same 
species, with $a_i$ the corresponding $s$-wave 
scattering length. The coupling between both 
components is governed by the inter-species 
interaction $g_{ab}\equiv g_{ba}$, which depends on the specific 
nature of the binary mixture. The total number of 
atoms in the mixture is $N=N_a+N_b$.

There are many experimental possibilities to study 
the dynamics of binary mixtures of BECs. We will 
restrict our study to one of them, which is experimentally 
feasible. We will consider binary mixtures made of 
$F=1$ $^{87}$Rb atoms populating the $m=\pm 1$ Zeeman 
sublevels~\cite{private}. This implementation greatly 
simplifies the dynamics as the inter- and intra-species 
couplings are very similar in magnitude. Of course this 
choice limits the phenomena which can be observed, e.g. the 
interesting symmetry breaking pattern discussed in 
Ref.~\cite{Satija2009}, which relies on the inter-species 
coupling being larger than the intra-species one, will
not take place, see Sec.~\ref{btm}.

On the other hand its simplicity allows to discuss 
in detail the different approaches taken in the 
literature, e.g. two-mode models of the GP 
equations, one dimensional reductions, etc. As 
occurred in the scalar case, the dynamical features 
contained in Eqs.~(\ref{GP-mixture}) can to a large 
extent be described by a simplified two-mode model 
for each component. In the next section we follow 
Refs.~\cite{Smerzi97},~\cite{Ananikian2006} 
and~\cite{Satija2009} and derive two-mode expressions 
for the scalar and binary case. The usual assumption 
of neglecting the overlaps 
involving the right and left modes gives rise 
to the so called standard two-mode (\sm) 
equations, while retaining them one also gets 
a closed system of equations, the improved two-mode 
(\im). Both the \sm and \im are also derived for the 
binary mixture case. 

\section{Two-mode approaches}
\label{sec:tw}

\begin{figure}[t]
\centering
\includegraphics[width=0.85\columnwidth,angle=0, clip=true]{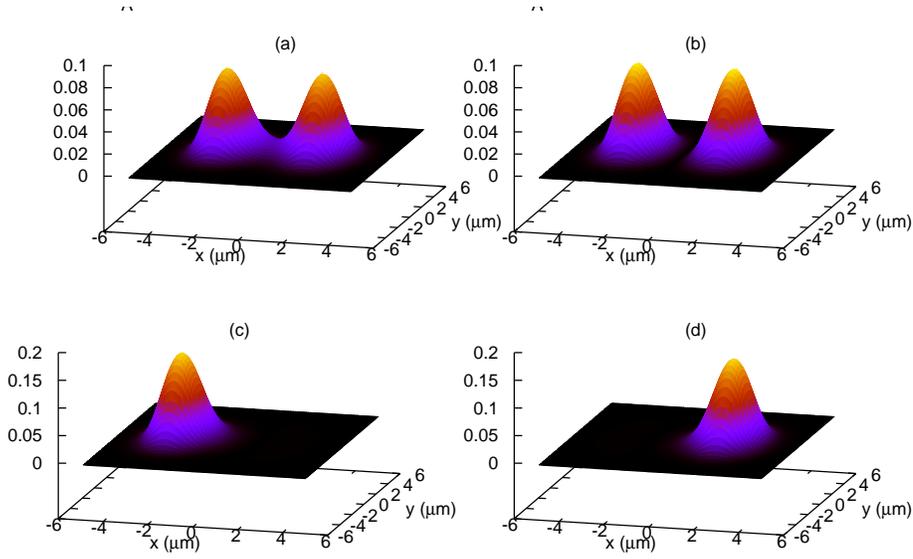}
\caption[]{3D depictions of the density, 
$\rho(x,y)=\int dz |\Psi(x,y,z)|^2$ in $(\mu m)^{-2}$, of the 
$(a)$ ground state, $(b)$ first excited state, 
$(c)$ left mode and $(d)$ right mode obtained 
by performing an imaginary time \tdgpetd calculation with 
the same conditions as in the experimental setup of 
Ref.~\cite{Albiez05}.} 
\label{local-modes}
\end{figure}

The two-mode approximation allows to study the 
dynamics of weakly linked Bose-Einstein condensates, 
without solving the full \tdgpetd nor reducing 
the dimensionality of the GP equation~\cite{Smerzi97,Ananikian2006}. 
Depending on the specific double-well potential, 
e.g. on the energy gap between the first two 
levels of the single particle Hamiltonian and the next two, 
it can provide an excellent description of 
the full GP solution. The 
relevant physical quantity is the ratio between the 
energy gap between the ground state and 
first excited state of the double-well potential, 
$\delta E_{0;1}=E_1-E_0$~\footnote{Note that this is zero if 
the barrier is infinitely high.} and the energy 
difference of the ground state and the 
second excited state, $\delta E_{0;2}=E_2-E_0$. The 
smaller the ratio $\delta E_{0;1}/\delta E_{0;2}$ the 
more accurate the two-mode approach. The two 
mode description characterizes the dynamics 
of the scalar condensate in a double-well 
potential with only two variables: the relative 
population and the phase difference between 
the left and right side of the potential barrier.

\subsection{Standard two-mode model for the single component case}
\label{2M-scalar} 

The GP equation for the scalar case corresponds to 
a particular limit of the GP equations for the 
binary mixture, Eqs.~(\ref{GP-mixture}), 
\begin{equation}
i\hbar{\partial \Psi({\bf r};t)\over \partial t} =
\left[-{\hbar^2\over 2 m}\nabla^2
+V({\bf r}) +g N|\Psi({\bf r};t)|^2 \right]\Psi({\bf r};t)\,.
\label{eq:sgp}
\end{equation}
We will make use of the following notation, 
$H_0=-(\hbar^2/ 2 m)\nabla^2+V({\bf r}) $, and 
$H_g[\Psi]=g N |\Psi({\bf r};t)|^2$. Let us recall 
the  two-mode approximation 
for a single component condensate in a double-well potential. We consider $N$ interacting 
atoms with atomic mass $m$, and coupling 
constant $g$, trapped in a symmetric double-well potential 
$V({\bf r})$. When both sides of the potential 
barrier are weakly linked, the total wave 
function can be approximately written as a superposition 
of two time-independent spatial wave functions 
$\Phi_{L(R)}({\bf r})$ mostly localized at the 
left (right) side of the trap:
\begin{equation}
\Psi({\bf r};t)= 
\Psi_{L}(t)\Phi_{L}({\bf r}) 
+ \Psi_{R}(t)\Phi_{R}({\bf r}) \,.
\label{scalar-2mode-ansatz}
\end{equation}
The left and right modes, can be expressed as 
linear combinations of the ground (+) and 
the first excited ($-$) states of the double-well 
potential including the interaction term. They 
satisfy, $(H_0+H_g[\Phi_\pm])\Phi_\pm=\mu_\pm \Phi_\pm$, and 
the left/right modes can be written as~\cite{Ananikian2006}:
\begin{eqnarray}
\Phi_{L}({\bf r})&=&
\frac{\Phi_{+}({\bf r}) +\Phi_{-}({\bf r})}{\sqrt{2}} \;,\qquad
\Phi_{R}({\bf r})=
\frac{\Phi_{+}({\bf r}) -\Phi_{-}({\bf r})}{\sqrt{2}}\,\,. 
\label{scalar-ansatz+-}
\end{eqnarray}
We observe that in a symmetric double-well, $\Phi_{\pm}$ have a well defined parity:
$
\Phi_{\pm}({\bf r})= \pm \Phi_{\pm}(-{\bf r})\,,
$
and therefore $ \langle \Phi_i \Phi_j \rangle 
= \delta_{ij}$ with $i,j=+,-$. Since they are 
stationary solutions of the GP equation, $\Phi_\pm$ are 
real functions, and so are the left and 
right modes $\Phi_{L(R)}$. The integrated density 
in the z-direction, $\rho(x,y)=\int dz |\Psi(x,y,z)|^2$, 
associated to the ground ($\Phi_+$), and first 
excited ($\Phi_-$) states are depicted in 
Fig.~\ref{local-modes} together with the densities 
associated to the left and right modes. The plots 
correspond to the experimental set up of Ref.~\cite{Albiez05}.

From the phase coherence properties of a BEC, 
one can assume that the wave function in each 
side of the trap has a well defined quantum 
phase $\phi_{j}(t)$, which is independent of 
the position but changes during the 
time evolution. We can write, 
\begin{equation}
\Psi_{j}(t)=\sqrt{N_{j}(t)} e^{i \phi_{j}(t)} \,,
\label{scalar-2mode-phase}
\end{equation}
where $N_{L(R)}(t)$ corresponds to the number 
of atoms on the left (right) side of the 
trap, and the total number of atoms is 
$N=N_{L}(t)+N_{R}(t)$. The weak link condition is 
fulfilled if $(\mu_--\mu_+)<<(1/2)(\mu_++\mu_-)$.

As a first step, we consider the so-called 
standard two-mode approximation (\sm), which 
neglects a certain set of 
overlapping integrals involving mixed products 
of $\Phi_{L}$ and $\Phi_{R}$. This approximation yields 
essentially the correct qualitative 
results in the scalar condensate although 
it may lead to incorrect quantitative 
predictions depending on the specific 
barrier properties~\cite{Albiez05,Ananikian2006} .

Inserting the two-mode ansatz 
(\ref{scalar-2mode-ansatz}) in the GP equation 
for a single component condensate (\ref{eq:sgp}) 
and neglecting terms involving mixed products 
of $\Phi_{L}$ and $\Phi_{R}$ of order larger than 
one, yields into a system of equations for the 
two localized modes which can be written 
in terms of two dynamical variables: the 
population imbalance $z(t)=[N_L(t)-N_R(t)]/N$ 
and the phase difference 
$\delta\phi(t)=\phi_R(t)-\phi_L(t)$ between 
each side of the barrier:
\begin{eqnarray}
\dot{z}(t)&=& - \omega_R\sqrt{1-z^2(t)} \,\sin {\delta\phi(t)}
\label{scalar-S2M-z2} \\
\dot{\delta\phi(t)}
&=&
\omega_R \,\Delta E + \omega_R {U_L+U_R\over 4 K} N z(t)
+\omega_R{z(t)\over \sqrt{1-z^2(t)}} \, 
\cos{\delta\phi(t)}\,,\nonumber
\end{eqnarray}
where $\omega_R=2 K/\hbar$ is the Rabi frequency 
and
\begin{eqnarray}
\Delta E &=& {E_L^0-E_R^0\over 2K} + {U_L - U_R \over 4K}N
\nonumber \\
E^0_{L(R)}&=& \int d {\bf r} 
\bigg[ 
{\hbar^2 \over 2m} \,
\big|\nabla\Phi_{L(R)}({\bf r})\big|^2  
+\Phi_{L(R)}^2({\bf r}) \,V({\bf r})\bigg]
\nonumber \\
K&=&  -\int d {\bf r} 
\bigg[ 
{\hbar^2 \over 2m} \,
\nabla\Phi_{L}({\bf r})\cdot \nabla\Phi_{R}({\bf r})
+\Phi_{L}({\bf r}) \,V({\bf r})\,\Phi_{R}({\bf r})\bigg]
\nonumber \\
U_{L(R)} &=&  
g\int\! d {\bf r} \,\Phi_{L(R)}^4({\bf r})
\;.
\label{scalar-parameters}
\end{eqnarray}
For a symmetric double-well, $E_L^0=E_R^0$ and $U_L=U_R\equiv U$, therefore
$\Delta E=0$.  
Moreover, the Rabi frequency only appears as a scale in 
the problem and thus can be absorbed in the 
time by rescaling $t\to \omega_R t$. Then, 
together with the definition 
$\Lambda\equiv N U /(\hbar \omega_R)$, we obtain, 
\begin{eqnarray}
\dot{z}(t)&=& - \sqrt{1-z^2(t)} \,\sin {\delta\phi(t)}
\label{scalar-S2M-z} \\
\dot{\delta\phi(t)}&=&{\Lambda  z(t)}
+{z(t)\over \sqrt{1-z^2(t)}} \, \cos{\delta\phi(t)}\,.\nonumber
\end{eqnarray}
Note that, $\Lambda>0$ and $\Lambda<0$ correspond to 
repulsive and attractive atom-atom interactions, 
respectively. 
There are different regimes depending on 
the initial values of the population imbalance 
and phase difference, $z(0)$ and $\delta\phi(0)$, 
Sec.~\ref{sec:regimes}.

From the energy functional of the GP equation 
(\ref{eq:sgp}):
\begin{equation}
E[\Psi({\bf r} ;t)]=\int d{\bf r} 
\bigg[{\hbar^2\over 2m}\Big|\vec{\nabla}
\Psi({\bf r} ;t) \Big|^2 + V({\bf r}) 
\big|\Psi({\bf r} ;t)\big|^2 
+ {g\over 2}\big|\Psi({\bf r} ;t)\big|^4 \bigg]
\end{equation}
and using the two-mode ansatz (\ref{scalar-2mode-ansatz}), 
we can define the conserved energy per particle of the 
system as:  
\begin{eqnarray}
H_{\rm} \equiv {E[\Psi({\bf r} ;t)] - C \over N K} = \Delta E\, z(t) + {U_L+U_R\over
  8K} N z^2(t) -
\sqrt{1-z^2(t)} \cos{\delta\phi(t)} \nonumber
\\
\end{eqnarray}
where $C$ is a rescaling constant. If we consider again a symmetric
double-well we have: 
\begin{equation}
H = {\Lambda\over 2} z^2(t) -
\sqrt{1-z^2(t)} \cos{\delta\phi(t)} \; . \label{energy-scalar-r}
\end{equation}
 Note that the 
equations of motion (\ref{scalar-S2M-z}) can be 
written in the Hamiltonian form:
\begin{equation}
\dot{z}= - {\partial H\over \partial \delta\phi} \;; \qquad 
\delta\dot{\phi} = {\partial H \over \partial z} 
\label{eq:mot}
\end{equation}
being $z$ and $\delta\phi$ canonical conjugate 
variables.

\subsection{Improved two-mode model for the single component case}
\label{sec:scaim}

Ananikian and Bergeman~\cite{Ananikian2006} 
noticed that for a symmetric double-well there was 
no need to neglect any of the overlapping 
integrals to obtain a closed set of equations 
relating $z$ and $\delta\phi$.

Thus, remaining in the two-mode approximation 
but retaining all the overlaps it is straightforward 
to write down the following set of equations 
(cf. Eqs.~(22) in Ref.~\cite{Ananikian2006}), 
called the ``improved two-mode'' (\im) equations, 
\begin{eqnarray}
\dot{z}(t)&=& - B \sqrt{1-z^2(t)} 
\,\sin {\delta\phi}(t) + C(1-z^2(t)) \sin 2 \delta\phi(t)
\label{scalar-I2M-z} \\
\dot{\delta\phi(t)}&=&{A  z(t)} +{B z(t)\over \sqrt{1-z^2(t)}} \,
\cos{\delta\phi}(t) - C z(t) \cos 2 \delta\phi(t)
\,.\nonumber
\end{eqnarray}
Defining, 
$\gamma_{ij}=g \int\! d {\bf r} \,\Phi_{i}^2({\bf r})
\Phi_{j}^2({\bf  r})\;,i,j=+,-$, we have, 
$A=N\left(10 \gamma_{+-}-\gamma_{++}-\gamma_{--}\right)/4$, 
$B=2 K+ N(\gamma_{--}-\gamma_{++})/2$, and  
$C=g N \int\! d {\bf r} \,\Phi_{L}^2({\bf r})\Phi_{R}^2({\bf r})$. 

As discussed in detail in Ref.~\cite{Ananikian2006},  
the physics arising from the \im is similar to the 
one present in the \sm. The \im, however, is in much 
better agreement with the \tdgpetd for a broader set 
of double-well potentials, as we will see in 
Sec.~\ref{sec:sc}. In particular, for the 
double-well considered in the experimental setup 
of the Heidelberg group~\cite{Albiez05} the \sm 
(with the corresponding microscopic parameters, 
$K$ and $U$, computed from the \tdgpetd with the 
experimental 3D potential and experimental coupling, 
$g$) does not correctly predict the physics of the 
experiment, mainly due to the importance of the 
neglected overlaps, and not to a dynamics far 
from a two-mode one.

\subsection{Regimes for the single component case}
\label{sec:regimes}

\begin{figure}[t]
\centering
\includegraphics[width=0.65\columnwidth,angle=0,clip=true]{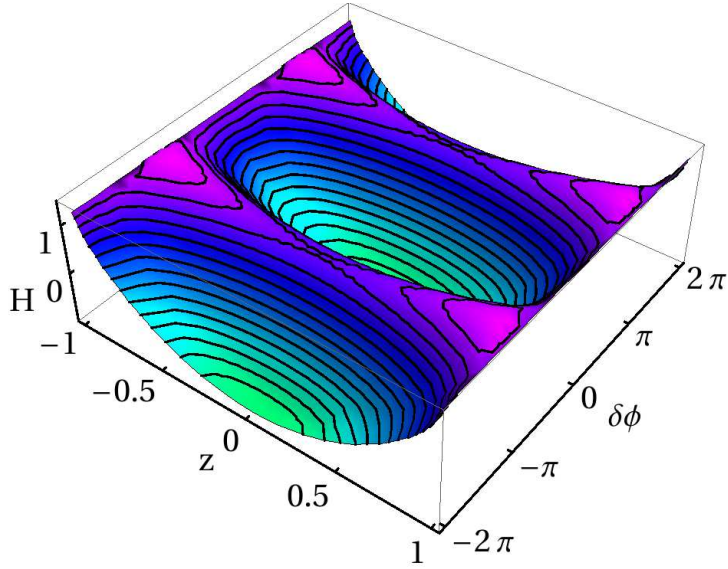}
 \caption[]{Energy surface, Eq.(\ref{energy-scalar-r}), for $\Lambda=2.5$. The
   lines on the surface correspond to possible trajectories of the system.}
\label{hamil}
\end{figure}

\subsubsection{Stability analysis} 
\noindent

In this section we use the \sm to analyze 
the stability of the single component system, and focus 
on the case of repulsive interactions $\Lambda>0$. 
Using the Hamiltonian (\ref{energy-scalar-r}) and 
the equations of motion (\ref{eq:mot}), 
the stationary points ($z^0$, $\delta\phi^0$) 
can be found by solving the equations: 
\begin{eqnarray}
\left. {\partial H \over \partial z}\right|_{z^0,\delta\phi^0} = 0 \; ;
\qquad  \left.{\partial H \over  \partial \delta\phi}\right|_{z^0,\delta\phi^0} = 0 \,.
\end{eqnarray}
To asses the stability of these points, we 
need to study the Hessian matrix of the 
system, which for the possible values of 
the phase difference,
$\delta\phi^0=0$ or $\pi$, is  
always diagonal and its eigenvalues are 
$\partial^2_{z}H |_{z^0,\delta\phi^0}$ and 
$\partial^2_{\delta\phi} H |_{z^0,\delta\phi^0}$. 
Depending on the sign of these eigenvalues 
the stationary points will be maxima, 
saddle points or minima. The stationary 
points and their stability are summarized 
in Table~\ref{tab-stat}. 

\begin{table}[b]
\centering
\begin{tabular}{r|c|c|c|c}
($z^0$, $\delta\phi^0$)  & stationary &   minimum & saddle & maximum    \\
\hline
($0$, $0$) & $\forall \Lambda $ & $\forall \Lambda$ &--- &---   \\
($0$, $\pi$)    & $\forall \Lambda$ & --- & $\Lambda > 1$ &$\Lambda < 1 $\\
($\pm\sqrt{1-1/\Lambda^2}$, $\pi$)   & $\Lambda>1$ & --- & --- & $\Lambda>1 $ \\
\end{tabular}
\caption{Stationary points of the system 
for repulsive interactions, $\Lambda > 0$, 
and their stability. \label{tab-stat}} 
\end{table}

The evolution of the system can be represented on 
a $z-\delta\phi$ plane, where the system follows 
trajectories with constant energy, $H$, see 
curves in Fig.~\ref{hamil}. Note that oscillations 
around a stationary point, closed curves, occur 
only if the central point is either a maximum or a 
minimum of the energy, but not a saddle point. As we 
will see in the following sections, these orbits 
will give rise to the Josephson oscillations and to the
zero- and $\pi$-modes.

\subsubsection{Symmetry between attractive and repulsive interactions}
\noindent

\begin{figure}[t]
$z(0)$
\begin{center}
\includegraphics[width=0.30\columnwidth, angle=0, clip=true]{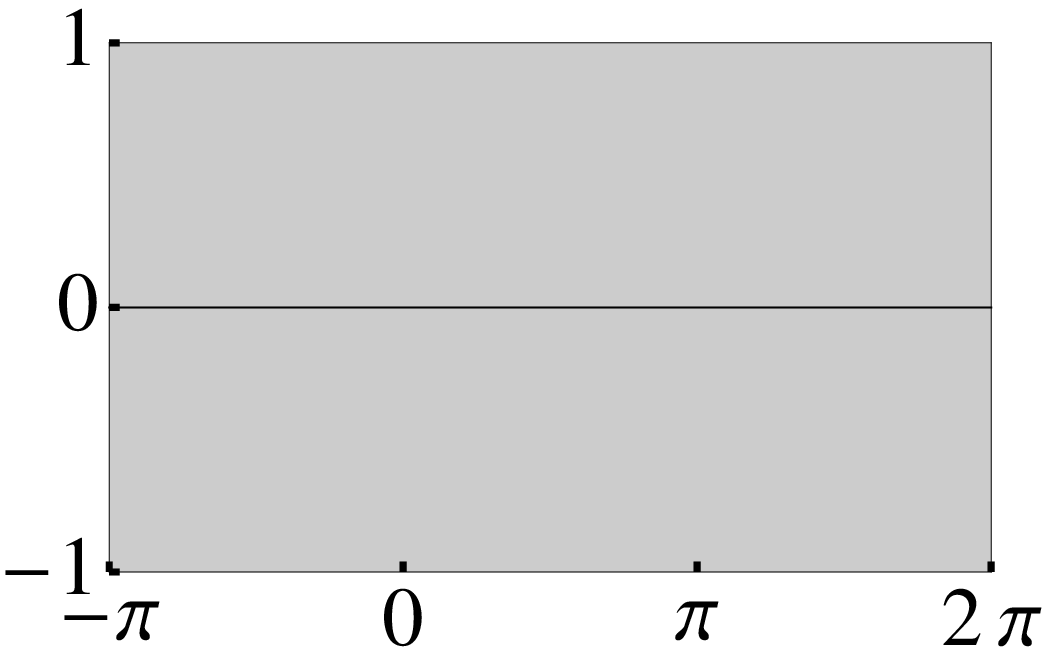}
\includegraphics[width=0.30\columnwidth, angle=0, clip=true]{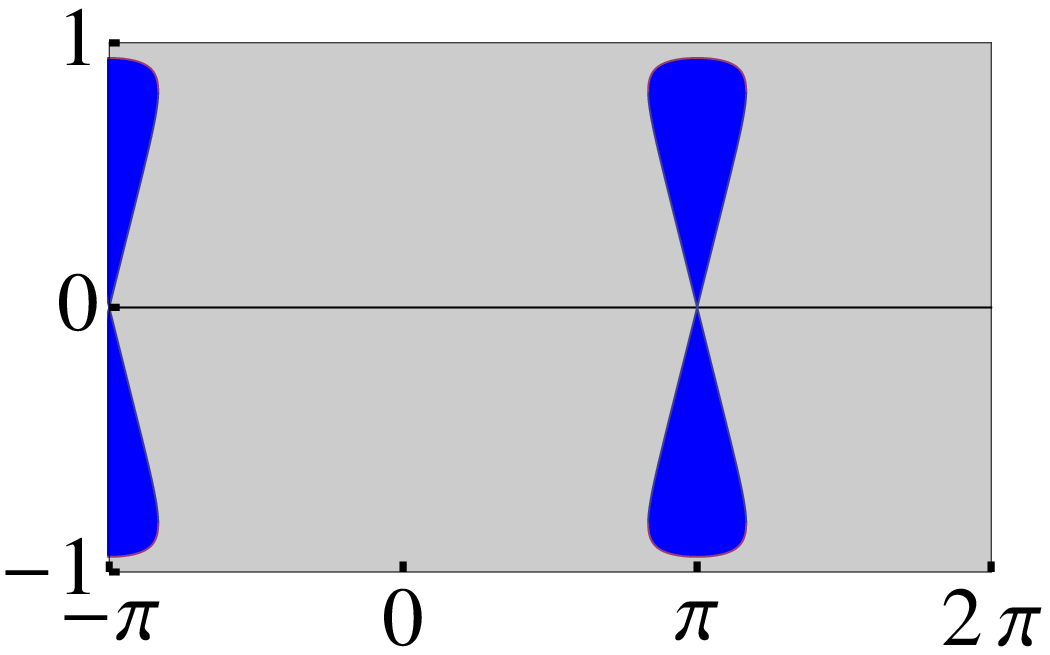}
\includegraphics[width=0.30\columnwidth, angle=0, clip=true]{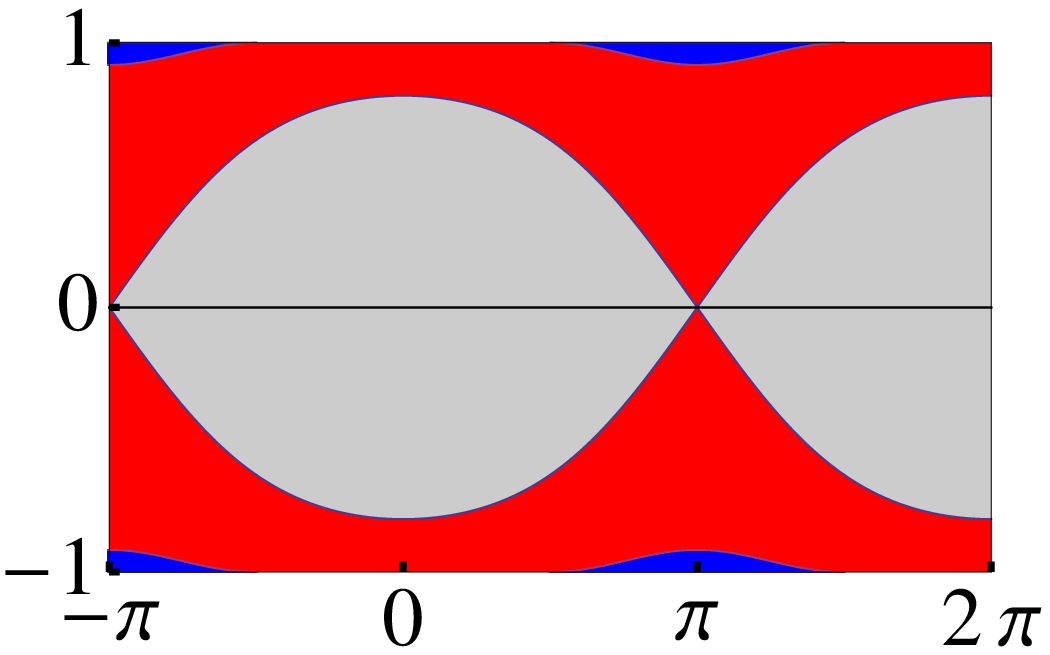}
\\
\includegraphics[width=0.30\columnwidth, angle=0, clip=true]{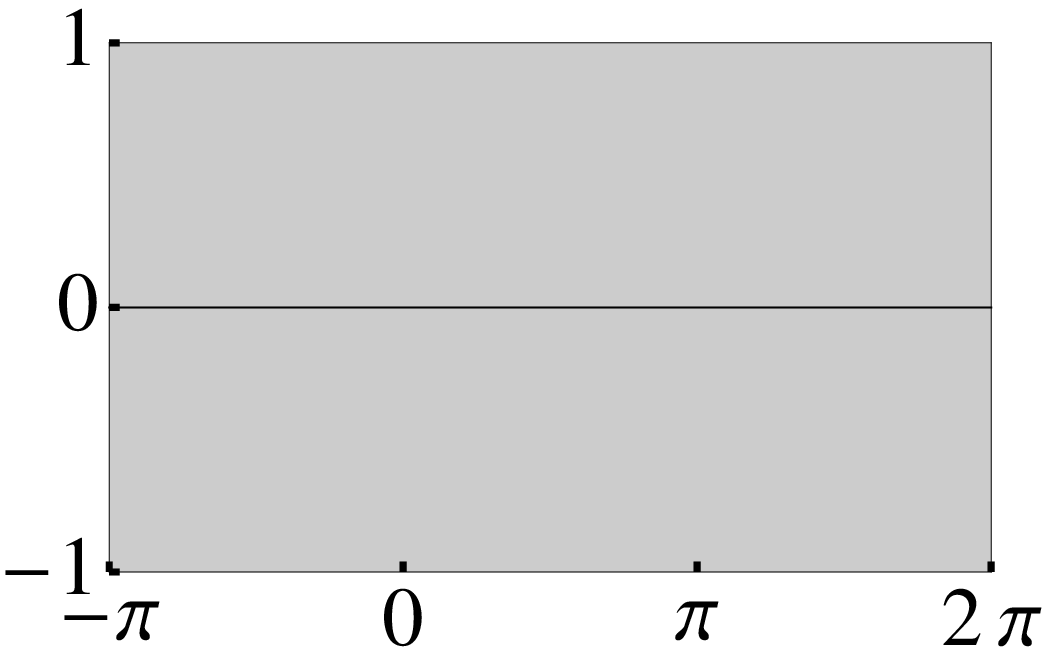}
\includegraphics[width=0.30\columnwidth, angle=0, clip=true]{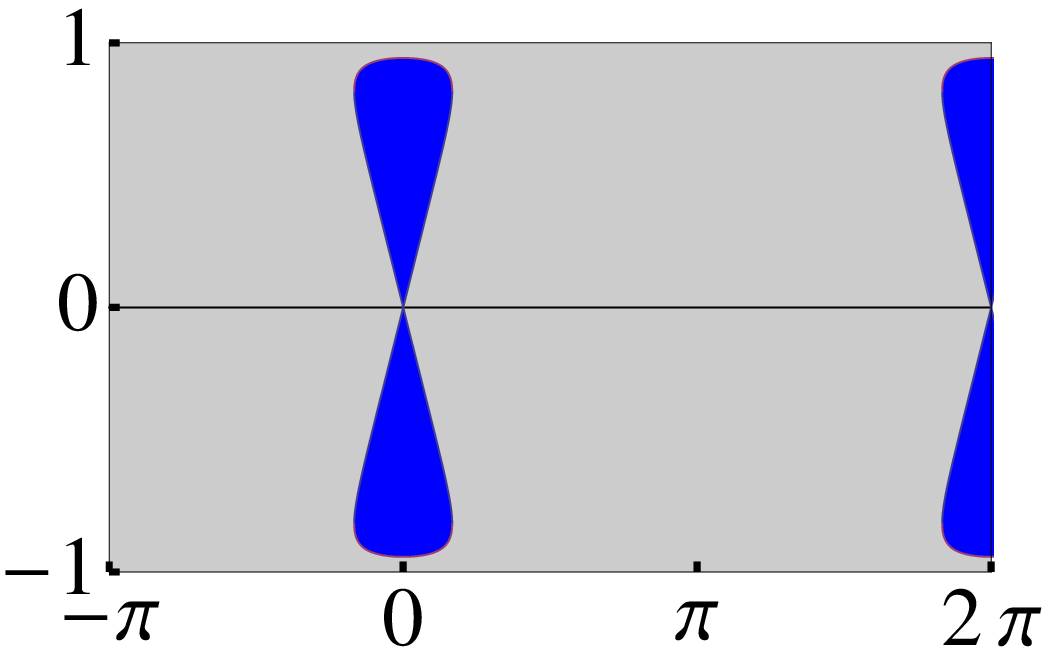}
\includegraphics[width=0.30\columnwidth, angle=0, clip=true]{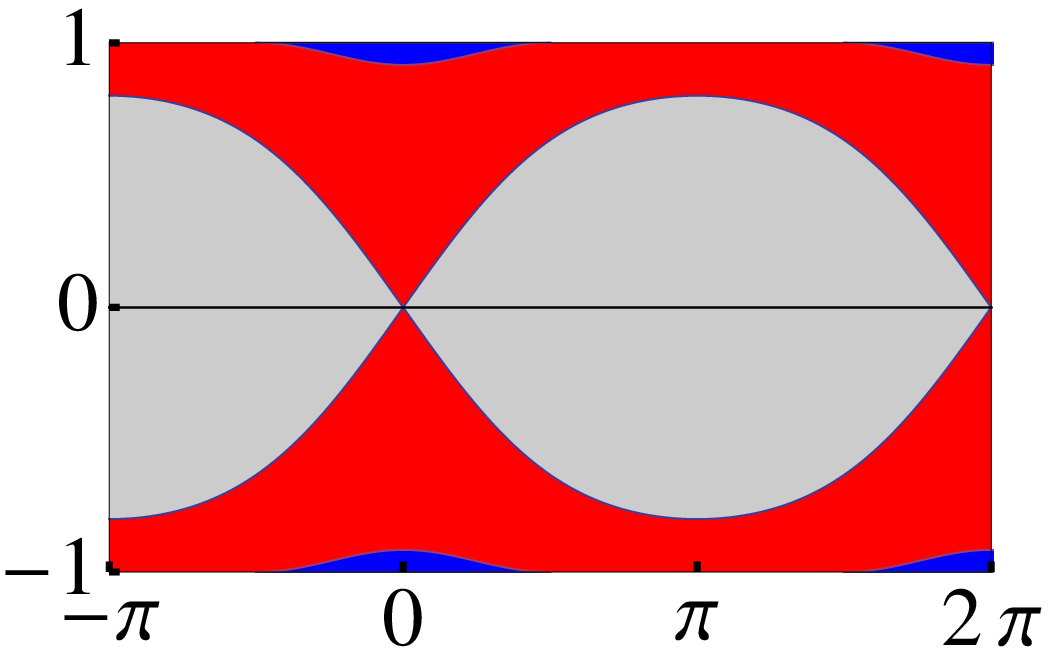}
\end{center}
\begin{flushright}
$\delta\phi(0)$
\end{flushright}
\begin{center}
\caption[]{Different regimes for a set of initial conditions, imbalance $z(0)$
  in the y-axis and phase difference $\delta\phi(0)$ in the x-axis. The upper
  panels correspond to repulsive interactions while the lower ones to
  attractive interactions. The values of $|\Lambda|$ are 0.5, 1.5, and 5.  
  for the left, middle and right panels respectively. Grey regions 
  correspond to Josephson oscillations, blue
  regions to $\pi$-modes (upper panels)  and zero-modes (lower panels), and
  red regions to running phase modes.} 
\label{fig:critic}
\end{center}
\end{figure}

The stability analysis has been presented 
only for repulsive interactions, but from 
the system (\ref{scalar-S2M-z}) we can see 
that if we change the interactions, 
$\Lambda\longrightarrow-\Lambda$, we recover 
the same system of equations if 
$\delta\phi \longrightarrow \pi-\delta\phi$:
\begin{eqnarray}
{d\over dt} z(t)&=& - \sqrt{1-z^2(t)} \,\sin{(\pi-\delta\phi(t))}
\\
{d\over dt} (\pi - \delta\phi(t))&=&-\Lambda  z(t)
- {z(t)\over \sqrt{1-z^2(t)}} \, \cos({\pi-\delta\phi(t)})\,,\nonumber
\end{eqnarray}
which means that the dynamics of the system 
and the different regimes are the same for 
both interactions, with a phase-shift of 
$\pi$. This can be seen in Fig.~\ref{fig:critic}, 
that shows the behavior of the system for a given 
set of initial conditions. The upper panels 
are for repulsive interactions $\Lambda>0$ and 
the lower ones for attractive interactions 
$\Lambda<0$. The grey regions correspond 
to Josephson oscillations, the blue regions 
to zero- and $\pi$-modes, and the red regions to 
running phase modes, as it will be seen in the following sections.

\subsubsection{Josephson dynamics} 
\noindent

\begin{figure}[t]
\begin{center}
\includegraphics[width=0.25\columnwidth, angle=0, clip=true]{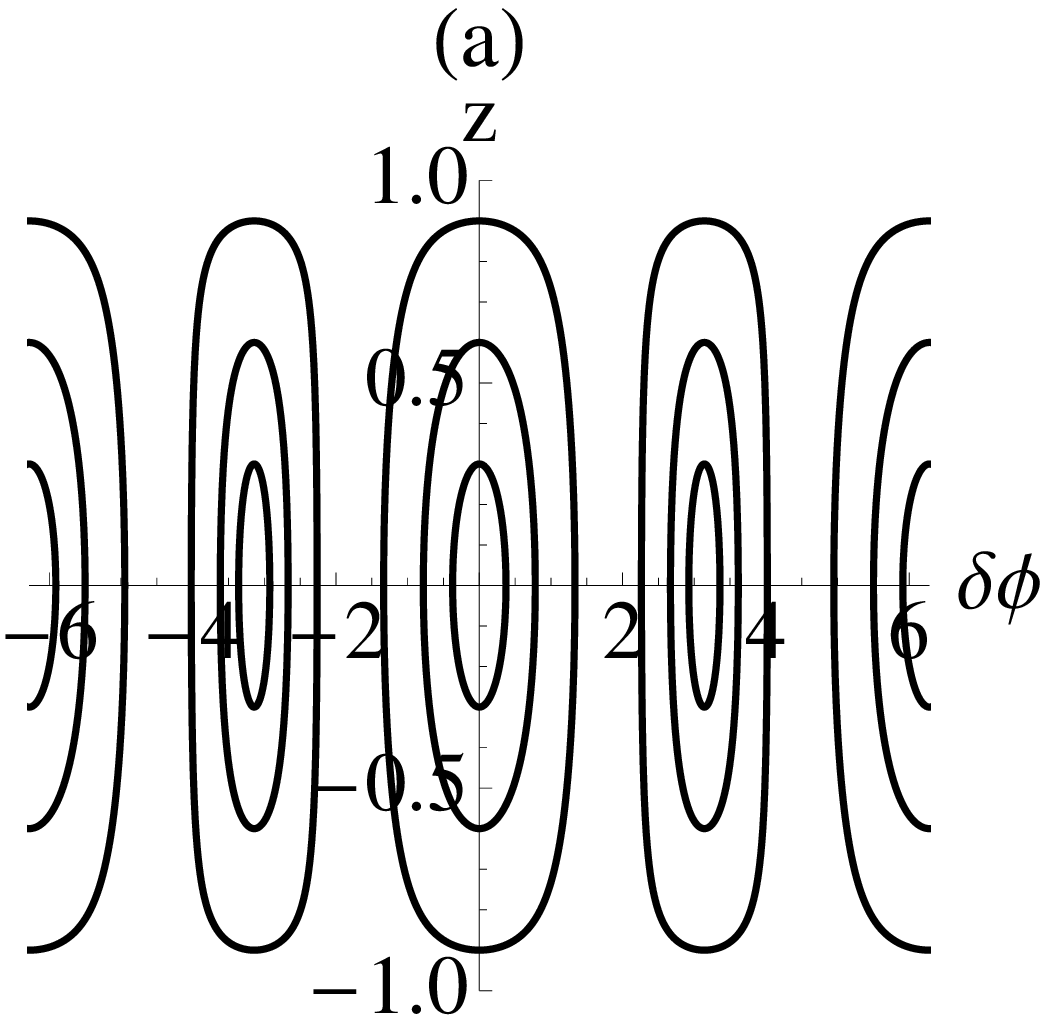}
\hspace{1cm}
\includegraphics[width=0.25\columnwidth, angle=0, clip=true]{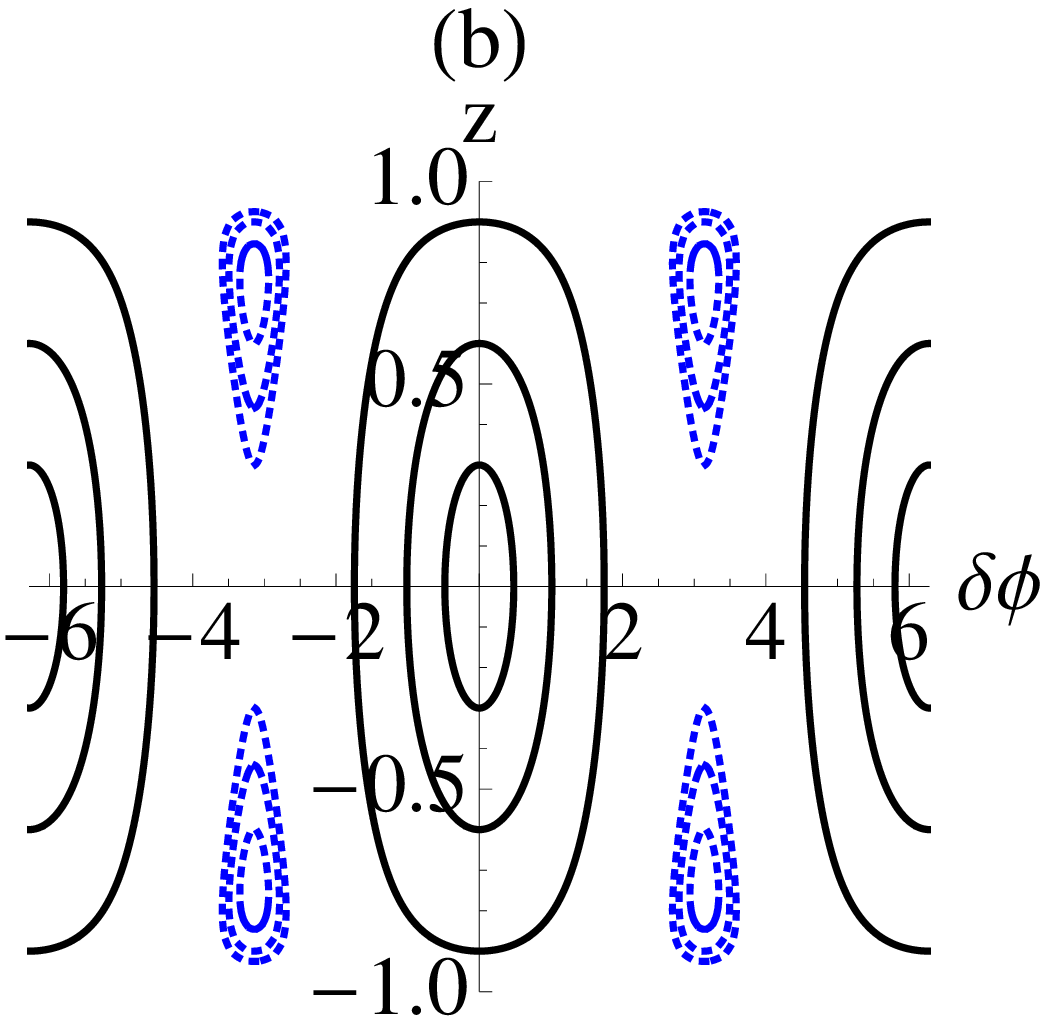}
\hspace{1cm}
\includegraphics[width=0.25\columnwidth, angle=0, clip=true]{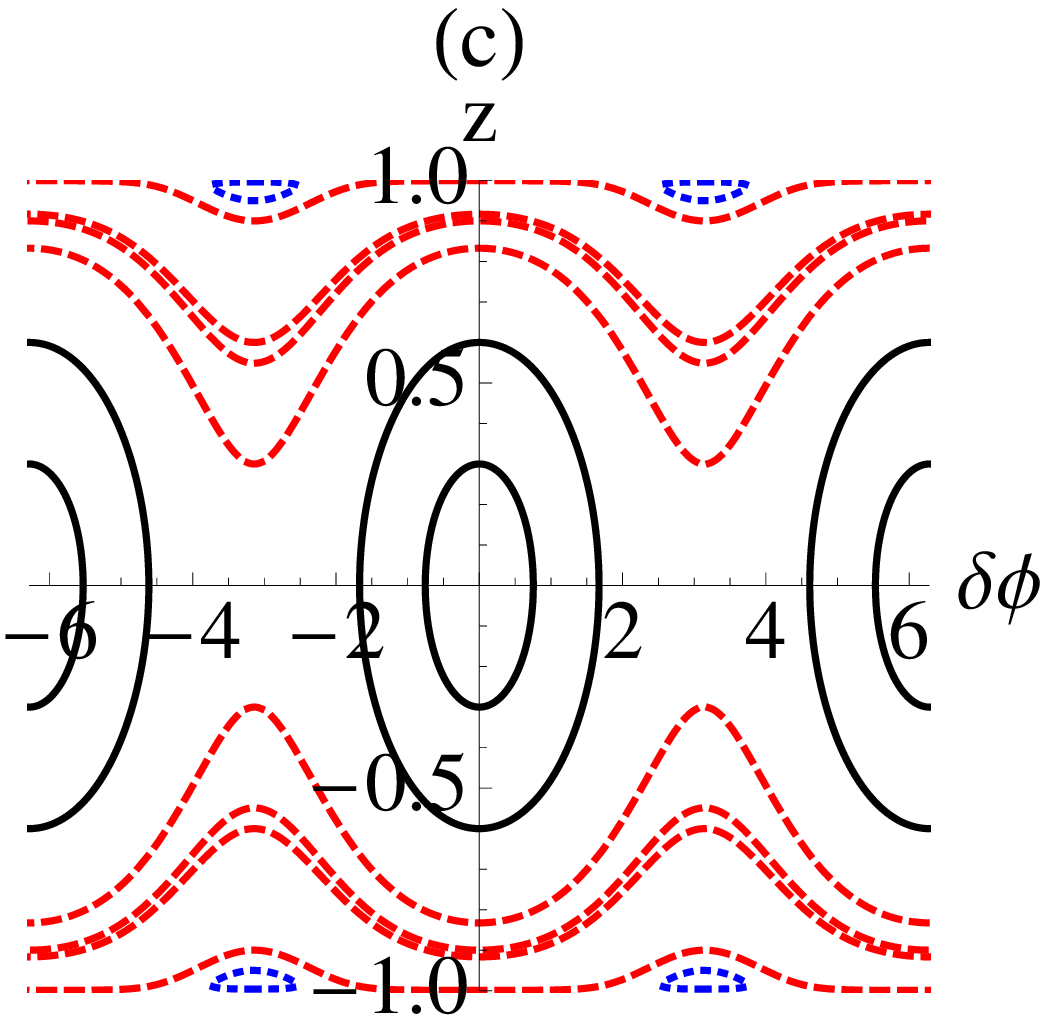}
\caption[]{$z-\phi$ representation of different constant 
energy trajectories for three values of $\Lambda$: 0.5 (a), 1.5 (b), and 
5 (c). Solid-black lines correspond to Josephson oscillations, 
dotted-blue to $\pi$-modes, and dashed-red
lines to running phase modes. 
\label{fig:tmscalar-r}}
\end{center}
\end{figure}

This regime is characterized by fast oscillating 
tunneling of population across the potential 
barrier. Plotted in a $z-\delta\phi$ map, the 
system evolves following closed trajectories around a 
minimum or a maximum ($z^0=0,\delta\phi^0$) configuration, 
with a zero time-average of the population imbalance, 
$<z>_t=0$. The stability analysis shows that for $\Lambda>-1$, 
which corresponds to repulsive or slightly 
attractive interactions, the stationary point 
$(z^0=0, \delta\phi^0=0$) is a minimum permitting 
Josephson oscillations around it. Analogously, 
for $\Lambda<1$, either attractive or slightly 
repulsive interactions, the stationary point 
($z^0=0$, $\delta\phi^0=\pi$) becomes a maximum, 
and therefore also allows for closed orbits 
around it. For $|\Lambda|>1$, there are Josephson 
oscillations around only one point: 
($z^0=0$, $\delta\phi^0=0$), or 
($z^0=0$, $\delta\phi^0=\pi$). However, in the region 
of weak interaction, $|\Lambda|<1$, the oscillations 
around both points are allowed. 

In panel (a) of Fig.~\ref{fig:tmscalar-r}, 
$\Lambda=0.5$, the black closed orbits 
around $\delta\phi^0=0$ or around 
$\delta\phi^0=\pi$ correspond to Josephson 
dynamics around these points. In panel 
(b) however, as $\Lambda=1.5>1$, only the 
origin can give rise to Josephson oscillations, 
so the closed orbits around $(z^0=0, \delta\phi^0=\pi)$ 
disappear. 

It is also interesting to study the behavior of the system 
for small oscillations around these two stationary 
points of zero imbalance, smallest orbits in 
Fig.~\ref{fig:tmscalar-r} (a). In this limit, the 
system (\ref{scalar-S2M-z}) can be linearized 
giving the dynamical equation:  
$\ddot{z}(t)= -z(t) (1+\Lambda\cos{\delta\phi^0})$ 
with $\cos{\delta\phi^0=\pm 1}$. The population 
imbalance performs sinusoidal oscillations with a frequency 
$\omega_J=\omega_R \sqrt{1+\Lambda\cos{\delta\phi^0}}$, independent 
of the initial conditions. Note that this frequency only exists when these
points are either maxima or minima. The phase 
difference oscillates with the same frequency but 
with a phase-shift of $\pi/2$ with respect to the imbalance. 
If the initial population imbalance increases, 
the dynamics of the system changes substantially 
to non-sinusoidal oscillations, and the 
frequency becomes dependent on the initial 
conditions.

\subsubsection{Macroscopic quantum self trapping} 
\noindent

In the case of repulsive interactions, we have seen that for $\Lambda>1$, the
stationary point ($z^0=0$, $\delta\phi^0=\pi$) becomes a saddle point and
there appear two maxima, ($z^0=\pm \sqrt{1-1/\Lambda^2}$, $\delta\phi^0=\pi$).
A similar behavior is found for attractive interactions. 
These stationary points allow for oscillations around them 
with $<z>_t\neq0$. In fact, in this regime, the imbalance has the same 
sign during the evolution, and therefore one of the wells is always 
overpopulated.

This regime is called macroscopic quantum self trapping, as the 
tunneling is strongly suppressed and the
particles remain mostly trapped in one of the wells. This is a 
phenomenon arising from the atom-atom interaction, which 
appears as a non-linearity in the Gross-Pitaevskii equation.      

The critical condition for the existence of the MQST regime can be found by
imposing that the system remains on one side of the trap ~\cite{Smerzi97}. For
a given set of initial conditions, $z(0)\neq 0$ and $\delta\phi(0)$, 
the system will remain trapped if,
\begin{eqnarray}
\Lambda&>& 2 \left({\sqrt{1-z(0)^2} \cos[\delta\phi(0)] +1 
\over z(0)^2}\right) \qquad {\rm for\;}\Lambda>1 \nonumber \\
\Lambda&<& 2 \left({\sqrt{1-z(0)^2} \cos[\delta\phi(0)] -1 
\over z(0)^2}\right) \qquad {\rm for\;}\Lambda<-1
\,, \label{MQST}
\end{eqnarray}
where the limits of the interaction parameter are due to the fact that only
when $|\Lambda|>1$ the $(z^0\neq 0,\delta\phi^0)$ stationary points exist. 

In this regime however, there are two different kind of MQST depending on 
whether the phase difference evolves bounded, giving the so-called 
zero- and $\pi$-modes, or whether it evolves unbounded, increasing 
(or decreasing) always in time,  giving rise to the running phase modes. 

For values of the interaction parameter of $1<|\Lambda|<2$ the only MQST
regime that one can have is the zero-mode for attractive interactions and
the $\pi$-mode for repulsive interactions (which are plotted in blue dotted
lines in panel (b) of Fig.~\ref{fig:tmscalar-r}). In these regimes the phase
difference evolves bounded around $\delta\phi=0$ and $\delta\phi=\pi$,
respectively.  
 
On the other hand, for values of $|\Lambda|>2$ one can have both classes of
MQST. In general however, for a given set of initial conditions, the system
will evolve following a running phase mode (dashed-red lines of panel (c) of
Fig.~\ref{fig:tmscalar-r}), because the values of
$z^0=\pm\sqrt{1-1/\Lambda^2}$, that allow closed orbits, are very close to 1
(see the small $\pi$-modes of panel (c) in blue dotted lines). 

In panel (c), one can see that the broadest closed orbit around 
($z^0\neq0$, $\delta\phi^0=\pi$), for $\Lambda>2$, is the 
one that goes through ($z=\pm1$, $\delta\phi=0$). Notice 
that an orbit that crosses the $\delta\phi=0$ axis in any other 
point, $z\neq \pm 1$, would correspond to a running phase mode. 
The case of attractive interactions can be understood 
by taking into account the phase-shift of $\pi$ in $\delta\phi$. 
The latter can be used to find the condition to 
have bounded or running phase difference modes. For a given set of initial
conditions ($z(0)$, $\delta\phi(0)$) fulfilling the self-trapping condition
(\ref{MQST}), the system will evolve in a bounded phase mode only if:
\begin{equation}
|\Lambda|<{2 \cos{\delta\phi(0)}\over \sqrt{1-z^2(0)}} \; .
\end{equation}

Moreover, in a zero- or a $\pi$-mode MQST, we can study small 
oscillations around the corresponding minima or maxima,
$z(t)=z^0+\delta z$ and  $\delta\phi(t)=\delta\phi^0+\hat{\delta\phi}(t)$, so
the linearized system (\ref{scalar-S2M-z}) becomes:
\begin{equation}
\delta \ddot{z}(t)= - \delta z(t) \left[1 + 
  \Lambda \cos{\delta\phi^0}{1-2 (z^0)^2\over\sqrt{1-(z^0)^2}}\right] 
\end{equation} 
which gives a sinusoidal behavior with a frequency:
\begin{equation}
\omega = \omega_R \sqrt{1+
  \Lambda \cos{\delta\phi^0}{1-2 (z^0)^2\over\sqrt{1-(z^0)^2}}} \; . 
\end{equation}

\subsection{Standard two-mode model for the binary mixture}
\label{S2M-sec}

Let us recall the two-mode approximation for weakly 
linked binary mixtures~\cite{ashab02,2component,xu2008,Satija2009}. 
The total wave function of each component is 
written as a superposition of two time-independent 
spatial wave functions localized in each well:
\begin{equation}
\Psi_j({\bf r};t)= \Psi_{j L}(t)\Phi_{j L}({\bf r}) + \Psi_{j
  R}(t)\Phi_{j R}({\bf r}) \,,
\label{2mode-ansatz}
\end{equation}
with 
$\langle \Phi_{i\alpha}|\Phi_{j\beta}\rangle=\delta_{ij}\delta_{\alpha\beta}$,
$i,j=a,b$ and $\alpha,\beta=L,R$. For a given 
component, the condensates in each side of the 
trap are weakly linked. Then, as in the scalar 
case, one can assume that the wave function in 
each side of the trap has a well defined quantum 
phase $\phi_{j,\alpha}(t)$, which is independent 
of the position but which changes during the 
time evolution. Thus,
\begin{equation}
\Psi_{j,\alpha}(t)
=
\sqrt{N_{j,\alpha}(t)} e^{i \phi_{j,\alpha}(t)}\,. 
\label{2mode-phase}
\end{equation}
$N_{j,L(R)}(t)$ corresponds to the population of 
the $j$-component on the left (right) side of 
the trap, with $N_j=N_{j,L}(t)+N_{j,R}(t)$. Inserting 
the two-mode ansatz (\ref{2mode-ansatz}) in the 
coupled GP equations for the mixture~(\ref{GP-mixture}), 
retaining up to first order crossed terms 
yields the following system of coupled equations:
\begin{eqnarray}
\dot{z_a}(t)&=& - {2K_a\over \hbar} \sqrt{1-z_a^2(t)}\sin {\delta\phi_a(t)}
\nonumber \\ 
\dot{\delta\phi_a}(t)&=& 
\Delta E_{a,b} 
+ {U_{aaL}+U_{aaR}\over 2\hbar} N_a z_a(t)
+ {U_{abL}+U_{abR}\over 2\hbar} N_b z_b(t)
\nonumber \\ &&  
+{2K_a \over \hbar} {z_a(t)\over
\sqrt{1-z_a^2(t)}}\cos{\delta\phi_a(t)}\nonumber \\
\dot{z_b}(t)&=& - {2K_b\over\hbar} \sqrt{1-z_b^2(t)}\sin {\delta\phi_b(t)}
\nonumber \\ 
\dot{\delta\phi_b}(t)&=&
\Delta E_{b,a}
+ {U_{bbL}+U_{bbR}\over 2\hbar} N_b z_b(t)
+ {U_{baL}+U_{baR}\over 2\hbar} N_a z_a(t)
\nonumber \\ &&  
+{2K_b\over\hbar} {z_b(t)\over \sqrt{1-z_b^2(t)}}\cos{\delta\phi_b(t)}\,.
\end{eqnarray}
where,
\begin{eqnarray}
\Delta E_{i,j} &=& {E_{iL}^0 - E_{iR}^0 \over \hbar} 
+ {U_{iiL}-U_{iiR}\over 2\hbar} N_i
+ {U_{ijL}-U_{ijR}\over 2\hbar} N_j  
\nonumber \\ 
E_{j\alpha}^0&=&
\int d{\bf r}\; 
\bigg[ 
{\hbar^2 \over 2m_j} |\nabla\Phi_{j\alpha}({\bf r})|^2+
\Phi_{j\alpha}^2 V({\bf r})
\bigg]\nonumber \\
K_{j}&=&-\int d{\bf r} \; 
\bigg[{
\hbar^2 \over 2m_j}  
\,\nabla\Phi_{jL}({\bf r})\cdot \nabla\Phi_{jR}({\bf r})
+\Phi_{jL}({\bf r}) \,V({\bf r})\,\Phi_{jR}({\bf r})
\bigg]
\nonumber \\
U_{ij \alpha}&=&g_{ij}\int d{\bf r}\; 
\Phi_{i \alpha}^2({\bf r})\Phi_{j \alpha}^2({\bf r})
\label{parameters}
\end{eqnarray}
with  $i,j=a,b\;{\rm and}\;\alpha=L,R$.
Let us consider a mixture with the same atomic 
mass for both components $M \equiv m_a=m_b$, 
which are trapped in the same symmetric 
double-well potential. Then, the localized 
modes are the same for both components but 
depend on the site: 
$\Phi_{L(R)}\equiv \Phi_{a,L(R)}=\Phi_{b,L(R)}$. 
Therefore, $E_{aL}^0=E_{bL}^0=E_{aR}^0=E_{bR}^0\equiv E$,
$U_{aaL}=U_{bbL}=U_{aaR}=U_{bbR}\equiv U$ and
$U_{abL}=U_{baL}=U_{abR}=U_{baR}\equiv \tilde{U}$, 
$K_a=K_b\equiv K$. Defining for each component 
the population imbalance and phase difference 
between both sides of the barrier, 
\begin{equation}
z_j(t)=(N_{j,L}(t)-N_{j,R}(t))/N_j \,,
\qquad
\delta\phi_j(t)=\phi_{j,R}(t)-\phi_{j,L}(t) 
\end{equation}
the above equations can be rewritten as:
\begin{eqnarray}
\dot{z_a}(t)&=& -\omega_R\sqrt{1-z_a^2(t)} \,\sin {\delta\phi_a(t)}
\label{S2M-z} \\
\dot{\delta\phi_a}(t)&=&{N_a U  z_a(t) 
+  N_b \tilde{U} z_b(t)\over\hbar}
+ \omega_R{z_a(t)\over \sqrt{1-z_a^2(t)}} \, \cos{\delta\phi_a}(t)\nonumber \\
\dot{z_b}(t)&=& -\omega_R \sqrt{1-z_b^2(t)} \, \sin {\delta\phi_b}(t) \nonumber \\
\dot{\delta\phi_b}(t)&=& 
{N_b U z_b(t) + N_a \tilde{U} z_a(t)\over\hbar} 
+\omega_R{z_b(t)\over \sqrt{1-z_b^2(t)}} \,\cos{\delta\phi_b}(t) 
\nonumber \,,
\end{eqnarray}
where $\omega_R=2 K/\hbar$ is the Rabi frequency, 
the same for both species. It is useful to define, 
$\Lambda=N U/\hbar \omega_R$, 
$\tilde{\Lambda}=N \tilde{U}/\hbar \omega_R$, 
$f_a=N_a/N$, $f_b=N_b/N$ and rescale the time 
as $t\to \omega_R t$, 
\begin{eqnarray}
\dot{z_a}(t)&=& - \sqrt{1-z_a^2(t)} \,\sin {\delta\phi_a}(t)
\label{S2M-z2} \\
\dot{\delta\phi_a}(t)&=&{f_a \Lambda z_a(t) +  f_b \tilde{\Lambda}
z_b(t)}
+{z_a(t)\over \sqrt{1-z_a^2(t)}} \, \cos{\delta\phi_a}(t)\nonumber \\
\dot{z_b}(t)&=& - \sqrt{1-z_b^2(t)} \, \sin {\delta\phi_b}(t) \nonumber \\
\dot{\delta\phi_b}(t)&=& {f_b \Lambda z_b(t) + f_a \tilde{\Lambda} z_a(t)} 
+{z_b(t)\over \sqrt{1-z_b^2(t)}} \,
\cos{\delta\phi_b}(t) \nonumber \,.
\end{eqnarray}
These equations correspond to two coupled 
nonrigid pendulums. The stability of these 
systems of equations have been analyzed 
recently in Ref.~\cite{xu2008}.

\subsection{Improved two-mode model for the binary mixture}

As was noted for the scalar case in 
Ref.~\cite{Ananikian2006}, it is not 
mandatory to neglect any of the overlaps 
to obtain a closed set of 
equations relating the population 
imbalances and phase differences for a symmetric double-well potential.
The complete set of two-mode equations were 
called the improved two-mode (\im) equations. 

In principle, if the experimental setup 
is appropriately chosen the left and right 
modes may be quite well localized at each 
side of the trap. In this case, the \sm 
equations are expected to provide quantitative 
agreement with the experimental data. When 
the two-modes are not so well localized, 
then it becomes necessary to consider 
the \im to have quantitative agreement. 
In~\cite{Ananikian2006} the authors considered 
explicitly the set up of the Heidelberg group 
and showed that the \im is necessary in 
the single component case to provide a quantitative 
understanding of the experimental data. 

Following similar steps as in the previous 
section and assuming the double-well potential 
to be symmetric as in the experiment, then 
the wave functions for the ground state and 
first excited state, $\Phi_{j\pm}({\bf r})$, 
have a well defined parity. The symmetry 
properties and the ortho-normalization conditions 
are capital to derive the coupled equations 
within the \im model:
$
\Phi_{j\pm}({\bf r})=\pm \Phi_{j \pm}(-{\bf r}) 
\label{I2M-wf}
$
,
$
\langle \Phi_{i\alpha}|\Phi_{j\beta}\rangle= \delta_{ij} \,
\delta_{\alpha\beta}\,, \,{\rm for\;} i,j=a,b \,\,\,{\rm and}\,\,\,
\alpha,\beta=+,- \,. 
$
The \im provides an exact description of the 
dynamics in the symmetric double-well potential, 
with no approximations beyond the assumption 
of a two-mode ansatz of the total wave function 
$\Psi_j({\bf r};t)$, Eq.~(\ref{2mode-ansatz}).

The resulting system of equations relating the 
population imbalance and phase difference for 
each component within the \im approximation 
reads\footnote{Our system of equations differs 
slightly with the previously derived ones, cf. 
appendix of Ref.~\cite{Satija2009}. We believe their 
system has some minor errors, which do not 
affect their discussion which is based on the 
\sm equations.}: 
\begin{eqnarray}
\dot{z}_a(t)&=&-{2K_{ab}\over \hbar}\sqrt{1-z_a^2(t)}\sin{\delta\phi_a}(t) \nonumber \\
\dot{\delta\phi}_a(t)&=&{\Delta_a(t)\over \hbar}
+{2K_{ab}(t)\over\hbar}{z_a(t)\over\sqrt{1-z_a^2(t)}} \cos{\delta\phi_a}(t)
\nonumber \\
\dot{z}_b(t)&=&-{2K_{ba}(t)\over\hbar}\sqrt{1-z_b^2(t)}\sin{\delta\phi_b}(t)
\nonumber \\
\dot{\delta\phi}_b(t)&=&{\Delta_b(t)\over \hbar}
+{2K_{ba}(t)\over\hbar}{z_b(t)\over\sqrt{1-z_b^2(t)}}\cos{\delta\phi_b}(t) \label{I2M}
\end{eqnarray}
with
\begin{eqnarray}
\Delta_a(t) &=&  2 \, \gamma_{+-}^{aa} N_a \,z_a(t)+ 2 \,
\gamma_{+-+-}^{aabb}\,N_b z_b(t)\,,
\nonumber \\
\Delta_b(t) &=& 2 \, \gamma_{+-}^{bb} \, N_b z_b(t) + 2 \,
\gamma_{+-+-}^{bbaa}\,N_a z_a(t) \,\,,
\end{eqnarray}
where we have defined
\begin{eqnarray}
\gamma_{\alpha\beta}^{i j}&=&g_{i j}\;\int d{\bf r}\,
\Phi_{i\alpha}^2({\bf r})\Phi_{j\beta}^2 ({\bf r})\label{gamma}\\
\gamma_{+-+-}^{aabb}&=& \gamma_{+-+-}^{bbaa}=g_{ab}\;\int d{\bf r}
\,  \Phi_{a+}({\bf r})\Phi_{a-}({\bf r})\Phi_{b+}({\bf r})\Phi_{b-}({\bf r})
\,,\nonumber 
\end{eqnarray}
and
\begin{eqnarray}
2 K_{ab}(t)&=& (\mu_-^a - \mu_+^a) +{1\over 2 } \Bigg[
N_a\left(\gamma_{++}^{aa}-\gamma_{--}^{aa}\right) \label{Kab-mu}  \\
& +& N_b\left(\gamma_{++}^{ab}- \gamma_{--}^{ab}-\gamma_{+-}^{ab}
+\gamma_{-+}^{ab} \right)\nonumber \\
& - & N_a
\Big(\gamma_{++}^{aa}+\gamma_{--}^{aa}-2\gamma_{+-}^{aa}\Big)
\sqrt{1-z_a^2(t)}\cos{\delta\phi_a}(t) \nonumber \\
& -&N_b
\Big(\gamma_{++}^{ab}+\gamma_{--}^{ab}-\gamma_{+-}^{ab}-\gamma_{-+}^{ab}\Big)
\sqrt{1-z_b^2(t)}\cos{\delta\phi_b(t)} \Bigg] \,. \nonumber
\end{eqnarray}
$\mu_+^j$ and $\mu_-^j$ are the chemical potentials 
of the ground and first excited state of the $j$ 
component, that can be calculated from the 
time-independent GP equation for $\Phi_{j\pm}$, 
respectively. Analogously one can define $2 K_{ba}$
by exchanging the subindex $a$ and $b$ in the 
previous expression.

Notice that we have kept the full 3D dependence 
of the wave functions $\Phi_{j\pm}({\bf r})$, 
instead of averaging the transverse spatial 
dependence as in Refs.~\cite{Ananikian2006,Satija2009}. 
Thus, the coupling parameters $g_{ij}$ in 
Eqs.~(\ref{gamma}) are the 3D ones and 
are not renormalized.

The equations for the \im are essentially 
similar to the \sm. The main difference is 
that the tunneling term, $K_{ab}(t)$, is 
time dependent and contains effects due to the 
interactions. As expected, if the localization 
of the modes is increased, i.e. by increasing 
the barrier height, $K_{ab}(t)$ approaches 
the constant value, $2 K_{ab}\to \mu_-^a - \mu_+^a$, 
which equals $2K$ of Eq.~(\ref{parameters}). 
The coupled equations obtained in the \im 
model reduce to well-known dynamical 
equations in two limiting cases:
\begin{itemize}
\item[i)] Setting to zero the overlapping 
integrals that involve mixed products of 
left and right modes of order larger than 1, 
the \im equations reduce to the \sm model for 
the mixture, Eqs.~(\ref{S2M-z}).
\item[ii)] Assuming a noninteracting mixture, 
the inter-species interaction is $g_{ab}=0$, 
and the \im equations for the mixture reduce 
to a two non-coupled system of equations, that are the dynamical 
equations of the \im for a single 
component, Sec.~\ref{sec:scaim}.
\end{itemize}

As discussed in the introduction we are interested 
in the particular case of a binary mixture made 
of atoms populating two different hyperfine states. 
Then, both components have the same mass $M$, and 
are trapped in the same symmetric double-well 
potential. We initially restrict to the case 
in which the inter-species interaction is also almost 
equal to the intra-species one, $g \equiv g_{aa}=g_{bb} \sim g_{ab}$. 
This is the situation for $F=1$ $m=\pm 1$ of $^{87}$Rb. This case allows 
straightforward comparisons between the results 
of the \im and the ones obtained by 
solving the \npse or \tdgpeod for a mixture explained 
in Sec.~\ref{3D-1D-sec}. 

The ground and first excited states in a symmetric 
double-well potential are the same for both 
components. Moreover, since $g=g_{ab}$ the overlap 
integrals (\ref{gamma}) reduce to:
\begin{eqnarray}
\gamma_{++}^{aa}&=&\gamma_{++}^{bb}=\gamma_{++}^{ab}\equiv\gamma_{++}  \nonumber \\
\gamma_{--}^{aa}&=&\gamma_{--}^{bb}=\gamma_{--}^{ab}\equiv\gamma_{--} \nonumber \\
\gamma_{+-}^{aa}&=&\gamma_{+-}^{bb}=\gamma_{+-}^{ab}=\gamma_{-+}^{ab}
=\gamma_{+-+-}^{aabb}\equiv\gamma_{+-} \,,
\label{gamma-same-particles} \end{eqnarray}
and the chemical potentials
$\mu_\alpha^a=\mu_\alpha^b \equiv \mu_\alpha$ with 
$\alpha=+,-$. This yields the following 
relations: $K_{ab}=K_{ba}$ and $\Delta_a=\Delta_b$. 
The \im system reduces to: 
\begin{eqnarray}
\dot{z}_a(t)&=&-{2K_{ab}(t)\over \hbar}\sqrt{1-z_a^2(t)}\sin{\delta\phi_a}(t) \nonumber \\
\dot{\delta\phi}_a(t)&=&{2 (N_a z_a(t)+N_b z_b(t))\gamma_{+-}\over
\hbar}+{2K_{ab}(t)\over
\hbar}{z_a(t)\over\sqrt{1-z_a^2(t)}}\cos{\delta\phi_a} \nonumber\\
\dot{z}_b(t)&=&-{2K_{ab}(t)\over
\hbar}\sqrt{1-z_b^2(t)}\sin{\delta\phi_b}
\nonumber \\
\dot{\delta\phi}_b(t)&=&{2 (N_a z_a(t)+N_b z_b(t))\gamma_{+-}\over
\hbar}+{2K_{ab}(t)\over
\hbar}{z_b(t)\over\sqrt{1-z_b^2(t)}}\cos{\delta\phi_b}(t)\,.
\end{eqnarray}
In this case both components obey the same 
system of coupled differential equations. Then, if 
the initial conditions are the same for both, 
$z_a(0)=z_b(0)$ and 
$\delta\phi_a (0)= \delta \phi_b (0)$, they will 
evolve with the same imbalance and phase, and 
no mixture effects will be observed. 

\subsection{Regimes for binary mixtures}
\label{sec:drbm}

We proceed now to analyze the stability of 
the system of equations~(\ref{S2M-z2}), cf. 
see the appendix of Ref.~\cite{ashab02}. 
As in the single component case, and in order to get analytical results that
allow for a physical insight, we perform the study in the framework of the \sm
approximation. First we note that an stationary 
point, defined by the equations: $\dot{z_i}=0$ 
and $\dot{\delta\phi}_i=0$, necessarily fulfills, 
\begin{eqnarray}
\sin\delta\phi_a=0 \quad \Rightarrow \delta\phi_a^0=0,\pi \nonumber \\
\sin\delta\phi_b=0 \quad \Rightarrow \delta\phi_b^0=0,\pi \,,
\label{eq:st1}
\end{eqnarray}
and the following system of equations, 
\begin{eqnarray}
z_a^0  = -z_b^0 \left({\Lambda\over \tilde\Lambda}+
{1 \over \tilde\Lambda f_b \sqrt{1-(z_b^0)^2}
  \cos\delta\phi_b^0}\right)\nonumber \\
z_b^0  = -z_a^0 \left({\Lambda\over \tilde\Lambda}+
{1 \over \tilde\Lambda f_a \sqrt{1-(z_a^0)^2} \cos\delta\phi_a^0}\right) \,.
\label{eq:st2}
\end{eqnarray}
Therefore there are four different cases: 
$(\delta\phi_a^0=0,\delta\phi_b^0=0)$, 
$(\delta\phi_a^0=0,\delta\phi_b^0=\pi)$, 
$(\delta\phi_a^0=\pi,\delta\phi_b^0=0)$, 
$(\delta\phi_a^0=\pi,\delta\phi_b^0=\pi)$, 
noting that in all of them there is an 
obvious stationary point, $z_a^0=z_b^0=0$. These 
stationary points will be referred to as 
``trivial stationary points''.  We need 
to find the conditions for non-trivial 
solutions in each case.  

The stability of the system is analyzed by 
considering small variations around the 
stationary points for each of the four 
situations. Defining the displacements 
$\eta_i$, 
\begin{eqnarray}
z_a(t) = z_a^0+\eta_a(t) \nonumber \\
z_b(t) = z_b^0+\eta_b(t) \,, 
\end{eqnarray}
the following system of equations 
for the $\eta$'s can be derived 
from Eqs.~(\ref{S2M-z2}) 
\begin{equation}
\left( \matrix{ \ddot{\eta_a} \cr
                \ddot{\eta_b} }
\right)
=
-\Omega^2 
\left( \matrix{ \eta_a \cr
                \eta_b }
\right)
\end{equation}
where, 
\begin{eqnarray}
\Omega^2 &=&  \omega_R^2 
\left(
\matrix{ 
 1 + (f_a \Lambda z_a^0+f_b \tilde\Lambda z_b^0)^2  & 0\cr
         0 & 
 1 + (f_a \tilde\Lambda z_a^0+ f_b \Lambda z_b^0)^2     \cr
}
\right) \nonumber \\
&+&
\omega_R^2 
\left(
\matrix{
f_a \Lambda \sqrt{1-(z_a^0)^2} \cos\delta\phi_a^0  & 
f_a \tilde\Lambda \sqrt{1-(z_a^0)^2} \cos\delta\phi_a^0  \cr 
f_b \tilde\Lambda\sqrt{1-(z_b^0)^2} \cos\delta\phi_b^0  & 
f_b \Lambda \sqrt{1-(z_b^0)^2} \cos\delta\phi_b^0   
}
\right) \,.
\label{eq:freq}
\end{eqnarray}
In Table~\ref{tabnm} we give the explicit values 
of the eigenfrequencies of $\Omega$ for the trivial 
stationary points, $z_i^0=0$. These are obtained 
for the case under consideration, where 
$\Lambda>0$ and $\tilde\Lambda>0$.  

Approximate simpler expressions for the same 
eigenfrequencies can be derived for the case 
when $\tilde\Lambda \sim \Lambda$. Defining 
$\tilde\Lambda=\Lambda(1+\beta)$ and retaining 
up to terms of order $\beta$, one obtains the 
frequencies listed in Table~\ref{tabnm2}.  

\begin{table}[t]
\centering
\begin{tabular}{l|l|l}
($\delta\phi_a^0,\delta\phi_b^0$)     & $\omega_1^2/\omega_R^2$   & $\omega_2^2/\omega_R^2$\\
\hline
(0,0) & 
$1+{\Lambda\over 2} \left(1 +\sqrt{(f_a-f_b)^2+4f_af_b(\tilde{\Lambda}/\Lambda)^2}\right)$
& 
$1+{\Lambda\over 2}\left(1-\sqrt{(f_a-f_b)^2+4f_af_b(\tilde{\Lambda}/\Lambda)^2}\right)$   \\
($\pi,\pi$)  & 
$1-{\Lambda\over 2}\left(1-\sqrt{(f_a-f_b)^2+4f_af_b(\tilde{\Lambda}/\Lambda)^2}\right)$
& 
$1-{\Lambda\over 2}\left(1+\sqrt{(f_a-f_b)^2+4f_af_b(\tilde{\Lambda}/\Lambda)^2}\right)$\\
(0,$\pi$)  & 
$1+{\Lambda\over 2} \left( (f_b-f_a)+\sqrt{1 - 4f_af_b(\tilde{\Lambda}/\Lambda)^2}\right)$
&$1+{\Lambda\over 2}\left( (f_b-f_a)-\sqrt{1 - 4f_af_b(\tilde{\Lambda}/\Lambda)^2}\right)$\\
($\pi$,0)   & 
$1+{\Lambda\over 2}\left( (f_a-f_b)+\sqrt{1 - 4f_af_b(\tilde{\Lambda}/\Lambda)^2}\right)$
& 
$1+{\Lambda\over 2}\left( (f_a-f_b)-\sqrt{1 - 4f_af_b(\tilde{\Lambda}/\Lambda)^2}\right)$\\
\end{tabular}
\caption{Square of the frequencies of the 
eigenmodes of the \sm system, 
Eqs.~(\ref{S2M-z2}), linearized around 
the trivial stationary points, $z_i^0=0$ 
for the four different $\delta\phi_i^0$ combinations.
\label{tabnm}}
\end{table}

\begin{table}
\centering
\begin{tabular}{l|l|l}
     ($\delta\phi_a,\delta\phi_b$)            & $\omega_1^2/\omega_R^2$   & $\omega_2^2/\omega_R^2$\\
\hline
 (0,0) & 
$1 + \Lambda (1 + 2 \beta f_a f_b) $
&
$1 - 2 \Lambda\beta f_a f_b  $   \\
 ($\pi,\pi$)   
&  $1 + 2 \Lambda\beta f_a f_b$
& $1 -\Lambda (1 + 2 \beta f_a f_b)$\\
($0,\pi$)
&$ 1- {2\beta f_af_b \Lambda \over f_a-f_b}$
&$ 1+(f_b-f_a)\Lambda+ {2\beta f_af_b \Lambda \over f_a-f_b}$\\
($\pi,0$)
&$ 1+(f_a-f_b)\Lambda- {2\beta f_af_b \Lambda \over f_a-f_b}$
&$ 1+ {2\beta f_af_b \Lambda \over f_a-f_b}$\\
\end{tabular}
\caption{Same as Table~\ref{tabnm} but retaining 
up to the first order in $\beta$, where 
$\tilde{\Lambda}=\Lambda(1+\beta)$. We 
assume $f_a> f_b$. \label{tabnm2}}
\end{table}

\subsubsection{Stationary points with $(\delta\phi_a^0=0,\delta\phi_b^0=0)$}
\noindent

In this case, the condition for the existence of 
non-trivial solutions to the equations (\ref{eq:st2}) 
depends on the slope at the origin of the two 
curves (\ref{eq:st2})~\cite{ashab02}. The condition 
\begin{equation}
\left( 
{ \Lambda\over \tilde\Lambda}+{1\over f_b \tilde\Lambda}
\right)
\left( 
{ \Lambda\over \tilde\Lambda}+{1\over f_a \tilde\Lambda}
\right) <1 \,,
\label{eq:c1}
\end{equation}
guarantees the existence of two additional solutions besides the trivial one.
If we restrict ourselves to the case of 
$\tilde\Lambda \sim\Lambda>0$, we have that (\ref{eq:c1}) 
cannot be fulfilled and 
therefore the only stationary point is 
the trivial one, $z_a^0=z_b^0=0$. 

In this case, it is straightforward to substitute the 
stationary point, $z_a^0=z_b^0=0$ and 
$\delta\phi_a^0=\delta\phi_b^0=0$
into Eq.~(\ref{eq:freq}) to get, 
\begin{equation}
\Omega^2= \omega_R^2 
\left(
\matrix{ 
 1  & 0\cr
 0 & 1       \cr
}
\right)
+
\omega_R^2 
\left(
\matrix{
f_a \Lambda    & f_a \tilde\Lambda   \cr 
f_b \tilde\Lambda  & f_b \Lambda   
}
\right)
\end{equation}
which has two eigenvalues, listed in Table~\ref{tabnm}.  

In the 
very polarized case, $f_a\sim 1, f_b\sim 0$, 
the population imbalance of the most populated 
component decouples from the less populated 
one and oscillates with the Josephson frequency 
$w_J=\omega_1$. The less populated component 
is driven by the other component and follows 
its dynamics, thus giving rise to ``anti-Josephson'' 
oscillations. The smaller frequency oscillation 
seen in the population imbalance of the less 
populated component is $\omega_2$, which in 
this case is very similar to $\omega_R$~\cite{Bruno2009}. 

Also interesting 
is the non-polarized case, $f_a=f_b=1/2$, 
then (assuming $\tilde\Lambda\sim \Lambda$, 
which is the case for $^{87}$Rb),    
\begin{eqnarray}
\ddot{z_a}(t)&=& -  \Lambda/2 (z_a(t)+z_b(t)) - z_a(t) \,,
\label{S2M-z-lin} \\
\ddot{z_b}(t)&=& - \Lambda/2 (z_a(t)+z_b(t))  - z_b(t) \nonumber \,.
\end{eqnarray} 
and defining $\Delta z(t)=z_a(t)+z_b(t)$, 
$\delta z(t)=z_a(t)-z_b(t)$ we have, 
\begin{eqnarray}
\ddot{\Delta z}(t)= -  (\Lambda+1) \Delta z(t) \;,\qquad 
\ddot{\delta z}(t)=   - \delta z(t) \nonumber \,.
\end{eqnarray} 
Therefore, $\Delta z$ behaves as the 
single component case, oscillating with 
the usual Josephson frequency, 
$w_J=\omega_R \sqrt{1+\Lambda}$ while 
$\delta z$ oscillates with the Rabi frequency, 
as would a single component case in the 
absence of atom-atom interactions. This mode can 
be further 
enhanced by imposing that $z_a(0)=-z_b(0)$ thus 
forcing both imbalances to oscillate with the 
same frequency. 

We have proposed in Ref.~\cite{Bruno2009} to use 
these two configurations to extract the frequencies 
governing the dynamics of the system in order to obtain 
the microscopic atom-atom interaction. 
The idea was to profit from the fact that the difference 
between the inter- and intra-species interaction 
is small for the case of $^{87}$Rb, 
$\tilde\Lambda=\Lambda(1+\beta)$, so we can use 
the expressions listed in Table~\ref{tabnm2}, 
$\omega_1^2 = \omega_R^2 (1 + \Lambda (1 + 2 \beta f_a f_b))$, and 
$\omega_2^2 = \omega_R^2 (1 - 2 \Lambda\beta f_a f_b)  $. Note 
that in the anti-Josephson case the oscillation with 
larger period is $\omega_2^2 = \omega_R^2 (1 + {\cal O}(\beta f_b))$ 
and the shorter is $\omega_1^2 = \omega_R^2(1 + \Lambda + {\cal O}(\beta
f_b))$, with $\beta<<1$ and $f_b<<1$, allowing to 
extract both the Rabi and Josephson frequencies 
with good precision. The second configuration only 
has one frequency which is 
$ \omega_1^2 = \omega_R^2(1 + \Lambda (1 + \beta/2 ))$ which allows 
to isolate the value of $\beta$.

\subsubsection{Stationary points with $(\delta\phi^0_a=\pi,\delta\phi^0_b=0)$}
\noindent

In this case, the condition for the existence 
of three stationary points is, 
\begin{equation}
\left( 
{ \Lambda\over \tilde\Lambda}-{1\over f_a \tilde\Lambda}
\right)
\left( 
{ \Lambda\over \tilde\Lambda}+{1\over f_b \tilde\Lambda}
\right) >1 \,.
\label{eq:c2}
\end{equation}
For the case considered here, $\tilde\Lambda \sim \Lambda$,
and, in most applications, $\Lambda>1$. Therefore, an appropriate choice 
of $f_a$ can ensure the existence of three stable points. 
The stability of the trivial solution is checked 
by studying,
\begin{eqnarray}
\Omega^2 &=&  \omega_R^2 
\left(
\matrix{ 
 1    & 0\cr
 0 &  1 \cr
}
\right) 
+
\omega_R^2 
\left(
\matrix{
f_a \Lambda   & 
f_a \tilde\Lambda   \cr 
-f_b \tilde\Lambda  & 
-f_b \Lambda    
}
\right) \,,
\end{eqnarray}
whose eigenvalues are listed in Table~\ref{tabnm}.
The stability of the other two solutions is easy 
to study with the same tools. Simple analytic 
expressions are only attainable for the case 
$\tilde\Lambda=\Lambda$. Then we have, 
\begin{eqnarray}
\Omega^2 &=&  \omega_R^2  
\left(
\matrix{ 
 1 + \Lambda (f_a  z_a^0+f_b  z_b^0)^2  & 0\cr
         0 & 
 1 + \Lambda (f_a z_a^0+ f_b z_b^0)^2     \cr
}
\right) \nonumber \\
&+&
\omega_R^2  \Lambda
\left(
\matrix{
f_a \sqrt{1-(z_a^0)^2}   & 
f_a  \sqrt{1-(z_a^0)^2}  \cr 
-f_b \sqrt{1-(z_b^0)^2}   & 
-f_b \sqrt{1-(z_b^0)^2}  
}
\right) \,,
\end{eqnarray}
whose eigenvalues are, 
\begin{equation}
\omega_1^2=\omega_R^2(\Lambda^2 (f_a z_a^0+f_b z_b^0)^2)\,,
\qquad
\omega_2^2=\omega_R^2(1+\Lambda^2 (f_a z_a^0+f_b z_b^0)^2) \, .
\end{equation}

\subsubsection{Stationary points with ($\delta\phi^0_a=\pi,\delta\phi^0_b=\pi$)} 
\noindent 

The condition for the existence of three 
stationary points is in this case~\cite{ashab02}, 
\begin{equation}
\left( 
{ \Lambda\over \tilde\Lambda}-{1\over f_b \tilde\Lambda}
\right)
\left( 
{ \Lambda\over \tilde\Lambda}-{1\over f_a \tilde\Lambda}
\right) <1 \,.
\label{eq:c3}
\end{equation}
The eigenvalues corresponding to small 
oscillations around the trivial point 
are listed in Table~\ref{tabnm}. Its 
dynamical stability depends on the specific 
values of $f_i$, $\tilde\Lambda$, $\Lambda$ and 
$\omega_R$. For the case $\tilde\Lambda=\Lambda$, 
it is stable provided that $\omega_R>\Lambda$. 

The eigenfrequencies for the non-trivial 
solution are the same as for the case 
$(\delta\phi_a=0,\delta\phi_b=\pi$). For 
the simplest case, $\tilde\Lambda=\Lambda$, 
they are, 
\begin{equation}
\omega_1^2=\omega_R^2(\Lambda^2 (f_a z_a^0+f_b z_b^0)^2)\,,
\qquad
\omega_2^2=\omega_R^2(1+\Lambda^2 (f_a z_a^0+f_b z_b^0)^2) \,.
\end{equation}

\section{Effective 1D mean field approaches}
\label{3D-1D-sec}

In the experimental realization~\cite{Albiez05} the 
condensate is confined by an asymmetric harmonic trap, 
characterized by $\omega_x, \omega_y$, and $\omega_z$, 
with a barrier on the $x$ direction. Thus, in a first 
approximation one can assume that the dynamics takes 
place mostly along the $x$ axis and derive descriptions 
of the system where the other two dimensions have been 
integrated out reducing the \tdgpetd equation to an effective 
1D equation. There are different procedures to derive 
effective one dimensional GP-like equations starting 
from the three dimensional one. Their generalization 
to binary mixtures, with two coupled GP equations, 
or spinor BEC, with three or more coupled GP equations, 
is presented below together with the single component case.

\subsection{One dimensional Gross-Pitaevskii-like equations (\gpod)}
\label{ssgp1d}

Assuming that most of the dynamics occurs in the direction 
which contains the barrier, the $x$ 
direction in our case, one can approximate the 
wave function of the system by
\begin{equation}
\Psi(x,y,z;t) \sim \Psi^{\rm 1D}(x;t) \;\varphi_{g.s.}(y) \;\varphi_{g.s.}(z)\,,
\end{equation}
where $\varphi_{g.s.}$ are the corresponding ground 
state wave functions for the trapping potential 
in the $y$ or $z$ direction in absence of 
interactions (in the case of harmonic traps 
they are Gaussian). In this way it can be 
shown~\cite{Oshanii98} that $\Psi^{\rm 1D}(x;t)$ 
fulfills a Gross-Pitaevskii-like 1D equation, 
\begin{equation}
i\hbar{\partial \Psi^{\rm 1D}(x;t)\over \partial t} 
=
\left[
-{\hbar^2\over 2 m}\partial^2_x
+V(x) 
+g_{\rm 1D} N |\Psi^{\rm 1D}(x;t)|^2 
\right]
\Psi^{\rm 1D}(x;t)\,,
\label{eq:gp1d}
\end{equation}
where the corresponding 1D coupling constant is 
obtained rescaling the 3D one, 
$g_{1D}=g/(2\pi a_{\perp}^2)$, with $a_{\perp}$ 
the transverse oscillator length, $a_{\perp}=\sqrt{\hbar /m \omega_\perp}$, 
with $\omega_\perp=\sqrt{\omega_z \omega_y}$. 

The extension to binary mixtures (and also 
to spinor condensates~\cite{Zhangyou}) may be 
written down readily, 
\begin{eqnarray}
&&i\hbar{\partial \Psi_a^{\rm 1D}(x;t)\over \partial t} =
\left[
-{\hbar^2\over 2 m}\partial^2_x
+V(x) + \sum_{j=a,b} g_{a\,j;\rm 1D} N_j|\Psi_j^{\rm 1D}(x;t)|^2 
\right]
\Psi^{\rm 1D}_a(x;t)
\nonumber \\
&&i\hbar{\partial \Psi_b^{\rm 1D}(x;t)\over \partial t} =
\left[
-{\hbar^2\over 2 m}\partial^2_x
+V(x) + \sum_{j=a,b} g_{b\,j;\rm 1D} N_j|\Psi_j^{\rm 1D}(x;t)|^2 
\right]
\Psi^{\rm 1D}_b(x;t)\,\nonumber \\
\label{eq:gp1dbm}
\end{eqnarray}
where, the rescaled couplings are 
$g_{ij; \rm 1D}=g_{ij}/(2\pi a_{\perp}^2)$.

\subsection{Non-polynomial Schr\"odinger equation (\npse)}

A more sophisticated reduction that includes to 
some extent the transverse motion of the 
elongated BEC in the corresponding potential 
is the so-called non-polynomial Schr\"odinger 
equation, proposed for a scalar BEC 
in Ref.~\cite{Salasnich2002}. The \npse recovers 
the previously discussed 1D reduction in 
the weakly interacting limit, but it has 
been shown to provide the best agreement with 
the experimental results on Josephson oscillations 
between two coupled BECs~\cite{AlbiezPhD}. 
The \npse for the scalar case reads,
\begin{eqnarray}
i\hbar{\partial \Psi (x;t) \over \partial t}& =&
\left[-{\hbar^2\over 2 m}\partial_x^2 +V(x) 
+g_{1D}  {N |\Psi(x;t)|^2\over 
\sqrt{1 + 2 a_s N|\Psi(x;t)|^2} } \right. \\
&
+
&
\left.
{\hbar \omega_\perp\over 2}
\left(
{1 \over 
\sqrt{1 + 2 a_s N|\Psi(x;t)|^2} }
+
\sqrt{1 + 2 a_s N|\Psi(x;t)|^2}
\right)
 \right]\Psi(x;t) \;.\nonumber 
\end{eqnarray}
The generalization of the \npse for two components 
in a binary mixture of BECs has been addressed in 
Ref.~\cite{Salasnich2006}. The system of equations, 
which become rather involved, can be greatly 
simplified in the case when all the interactions, 
both intra- and inter-species, are equal:
\begin{eqnarray}
i\hbar{\partial \Psi_j(x;t)\over \partial t} &=&
\left[-{\hbar^2\over 2 m_j}\partial_x^2 +V 
+g_{1D} { \rho(x;t)\over 
\sqrt{1 + 2 a_s  \rho(x;t) } }\right.\nonumber \\
&
+
&\left.
{\hbar \omega_\perp\over 2}
\left(
{1 \over 
\sqrt{1 + 2 a_s\rho(x;t)} }
+ 
\sqrt{1 + 2 a_s\rho(x;t)}
\right)
 \right]\Psi_j(x;t)
\label{npse}
\end{eqnarray}
where $\rho(x;t)= N_a |\Psi_a(x;t)|^2 + N_b |\Psi_b(x;t)|^2$, 
$j=a,b$, and, as before, $g_{1D}=g/(2\pi a_{\perp}^2)$, 
$g\equiv g_{aa}=g_{bb}=g_{ab}=g_{ba}$, and  
$\int dx |\Psi_j(x)|^2=1\,$.

\section{Numerical solutions of the 3D Gross-Pitaevskii equation: single component}
\label{sec:sc}

\begin{figure}[t]
\centering
\includegraphics[width=.8\columnwidth,angle=0, clip=true]{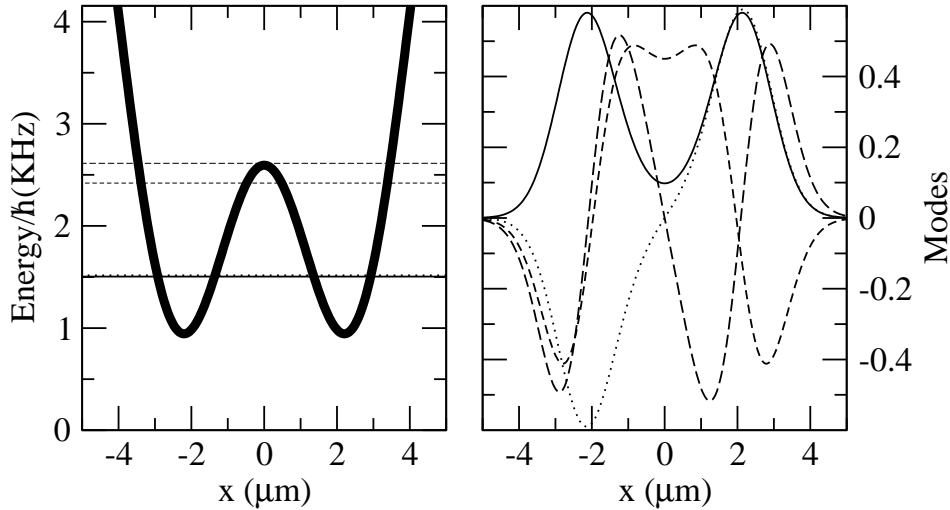}
\caption[]{ (left) Depiction of the potential 
in the $x$ direction in units of $\hbar$. The 
horizontal lines correspond to the single 
particle eigenenergies of the single particle 
Hamiltonian. (right) The first four single 
particle modes corresponding to the energies 
depicted in the left. 
\label{fig:potential}}
\end{figure}

Before analyzing the binary mixtures 
in the next section, we will present here numerical 
results for the single component to illustrate 
the main differences between the various two-mode 
models and 1D reductions. 

As discussed in the introduction, we consider the 
same setup and the same trap parameters as in 
the experiments of the Heidelberg group~\cite{Albiez05}. 
There, a condensate of $^{87}$Rb with $1150$ atoms 
is confined to a fairly small region of 
$\sim 5\, \mu$m through the potential, 
\begin{eqnarray}
V({\bf r})&=&
{1\over 2} M(\omega_x^2 x^2 +\omega_y^2 y^2 +\omega_z^2 z^2)
+V_0 \cos^2(\pi x /q_0 )
\label{potential}
\end{eqnarray}
with 
$\omega_x=2 \pi \times 78$ Hz, 
$\omega_y=2 \pi \times 66$ Hz,
$\omega_z=2 \pi \times 90$ Hz, 
$q_0=5.2 \mu$m, and $V_0=413 \,h$ Hz. In 
Fig.~\ref{fig:potential} we show the potential 
in the $x$ direction together with the first 
four energy levels of the single particle 
Hamiltonian and the corresponding modes. 
The energy levels of the single particle 
Hamiltonian show a clear separation between 
the two first eigenvalues, ground and 
first excited state, which are almost degenerate, 
and the next two. 

The atom-atom interaction strength is in 
this case, $g=4\pi\hbar^2 a /M$. 
The scattering length for $^{87}$Rb is 
$a=100.87 a_B$, therefore $g/\hbar$=0.04878 KHz $\mu m^3$. 
Noting that the number 
of atoms is known up to 10\% in the experiment, 
the relevant product, $gN/\hbar$ is in the range  
$[51.22,60.98]$ KHz $\mu m^3$. 
Ref.~\cite{Ananikian2006} uses 
a value of 58.8 KHz $\mu m^3$ to simulate the experimental 
setup. This large value of $gN$ corresponds to 
a situation similar to panel (c) of 
Fig.~\ref{fig:tmscalar-r}, where the possible 
dynamical situations we can have are: 
Josephson oscillations, i.e. closed orbits 
around the stationary point $(z^0,\delta\phi^0)=(0,0)$, 
and self trapping regimes, usually funning phase modes.

In the experiments, the system is prepared in a 
slightly uneven double-well potential 
which produces an initial population imbalance 
between both sides of the barrier. At $t=0$ the 
asymmetry is removed and the BEC is left to evolve 
in a symmetric double-well potential. In our 
numerical simulations the initial states with 
either $\delta\phi(0)=0$ or $\pi$ are constructed 
in a different way than in the experiment. 
We build initial states which are by 
construction two-mode-like. First, we obtain 
numerically the ground and first excited states 
of the condensate in the double-well potential 
by solving the time independent GP equation (both for 
the 1D reductions and the 3D case), then use 
those to build the left and right modes, 
Eq. (\ref{scalar-ansatz+-}), and 
finally construct initial states of any given 
initial imbalance, $z_0$: 
$\Psi_{z_0}({\bf r};t=0)= \alpha \phi_L({\bf r}) 
+ e^{l\imath \pi}\beta\phi_R({\bf r})$, 
with $\alpha^2+\beta^2=1$ and $\alpha^2-\beta^2=z_0$. 
The ground and first excited states are obtained 
by a standard imaginary time evolution of the 
equation from an initial state with the proper 
parity. The density profiles of the ground, 
first excited and left and right modes 
computed numerically are plotted in 
Fig.~\ref{local-modes}. As can be seen, the 
left/right modes are indeed well localized 
at each side of the barrier.

From these ground and first excited states 
we compute all the parameters entering in the 
\sm and \im descriptions presented in 
Secs.~\ref{2M-scalar}, and~\ref{sec:scaim}. 
The actual values of the parameters are, 
$K/\hbar= 0.00799$ KHz, and 
$N U/\hbar=1.19841$ KHz for the \sm and 
$A/\hbar=1.19372$ KHz, $B/\hbar=0.03683$ KHz, 
and $C/\hbar=0.0023590$ KHz for the \im
\footnote{These values compare reasonably well with 
the ones provided in page 33 of Albiez PhD 
thesis~\cite{AlbiezPhD}, there they are 
given in units of $\omega_x$: $A/\omega_x= 2.43572$, 
$B/\omega_x=0.0751497$, $C/\omega_x=0.0048$, 
and $K/\omega_x=0.0163$.}.  
The values of the overlaps are:  
$N \gamma_{++}/\hbar=0.581746$ KHz, 
$N \gamma_{+-}/\hbar=0.59803$ KHz and 
$N \gamma_{--}/\hbar=0.623769$ KHz. These numbers 
are used to generate the comparisons to \sm or \im in 
the following figures. 

In the full \tdgpetd simulations we define 
the number of atoms in the left well as: 
$N_L(t)=
\int_{-\infty}^{0} dx  
\int_{-\infty}^{\infty} dy 
\int_{-\infty}^{\infty} dz \,\, 
|\Psi({\bf r};t)|^2 \,.$
The number of atoms in the right well is computed 
as $N_R(t)=N-N_L(t)$. From these values, the population 
imbalance reads, $z(t)=(N_L(t)-N_R(t))/N$. Analogous 
definitions are used in the \tdgpeod and \npse equations. 

The phase difference between both 
sides of the potential barrier is computed in 
the following way. The phase at each point at a certain 
time, $\phi(x,y,z;t)$, is:
\begin{equation}
\Psi(x,y,z;t)= \sqrt{\rho(x,y,z;t)} \exp(\imath \;\phi(x,y,z;t))\,,
\label{bigphase}
\end{equation}
where the local density, $\rho(x,y,z;t)=|\Psi(x,y,z;t)|^2\,$.

Averaged densities are defined as, i.e. integrating over 
the $z$ component, 
\begin{equation}
\rho(x,y;t)=\int_{-\infty}^{\infty} dz \,\, \rho(x,y,z;t)\,. 
\end{equation}

To visualize the phase coherence along some of the 
planes we define, e.g. integrating the $z$ component, 
\begin{equation}
\phi(x,y;t)={1\over \rho(x,y;t)} 
\int_{-\infty}^{\infty} dz \,\, \rho(x,y,z;t) \;\phi(x,y,z;t) \,.
\end{equation}
The phase on the left, $\phi_L(t)$, is defined as, 
\begin{equation}
\phi_{L}(t)={1\over N_L(t)}
\int_{-\infty}^{0} dx  
\int_{-\infty}^{\infty} dy 
\int_{-\infty}^{\infty} dz \,\, 
\rho(x,y,z;t) \;\phi(x,y,z;t) \,.
\end{equation}
The phase on the right is defined accordingly.

\begin{figure}[t]
\centering
\includegraphics[width=1.\columnwidth,angle=0, clip=true]{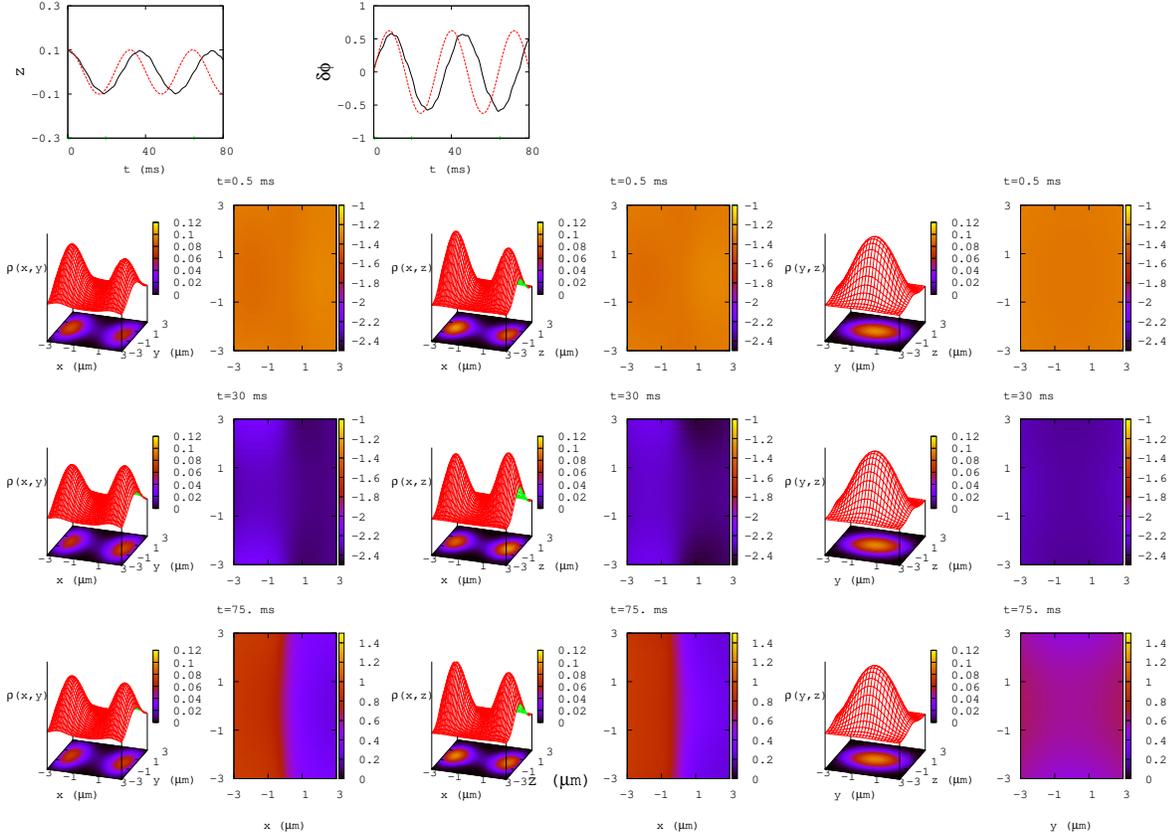}
\caption[]{The two smaller plots above depict 
in solid-black line the \tdgpetd time evolution of $z$ 
(left) and $\delta\phi$ (right), 
computed as explained in the text compared to the 
\im predictions in dashed-red. Then we show 3D pictures 
complemented with contour plots, left, of $\rho(x,y;t)$, 
$\rho(x,z;t)$ and $\rho(y,z;t)$ at three different times, 
0.5 ms (upper), 30 ms (middle) 
and 75 ms (lower), respectively. On the right of each plot 
we present a contour plot of the averaged quantum phase 
$\phi(x,y;t)$, $\phi(x,z;t)$ and $\phi(y,z;t)$ at the same times. They correspond to the 
first run presented in Fig.~\ref{comp}, $z(0)=0.1$ and 
$\delta\phi(0)=0$. 
\label{phase3d}}
\end{figure}

The way to implement the above averages 
over the phase has been done in the following way, 
\begin{eqnarray}
\phi(x,y;t) &=&{\rm arctan} 
{\int_{-\infty}^\infty dz \;{\rm Im}[\Psi(x,y,z;t)] \;\rho(x,y,z;t) \over 
\int_{-\infty}^\infty dz \;{\rm Re}[\Psi(x,y,z;t)] \;\rho(x,y,z;t) } \,, \nonumber\\
\phi_L(t) &=&{\rm arctan} 
{\int_{-\infty}^0 dx \int_{-\infty}^\infty dy \int_{-\infty}^\infty dz 
\;{\rm Im}[\Psi(x,y,z;t)] \;\rho(x,y,z;t) \over 
\int_{-\infty}^0 dx \int_{-\infty}^\infty dy \int_{-\infty}^\infty dz 
 \;{\rm Re}[\Psi(x,y,z;t)] \;\rho(x,y,z;t) } \;. \nonumber\\
\end{eqnarray}

\subsection{\tdgpetd results}

\begin{figure}[t]
\centering
\includegraphics[width=0.45\columnwidth,angle=0, clip=true]{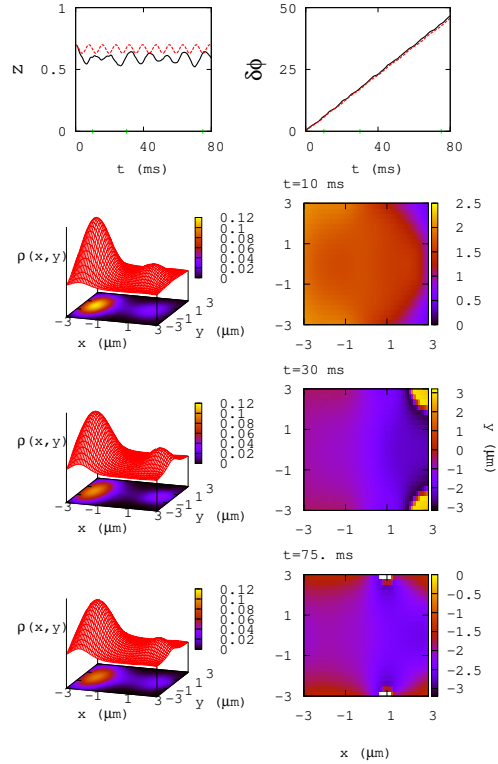}
\caption[]{Similar to Fig.~\ref{phase3d} but for a self-trapped case, 
$z(0)=0.7$, $\delta\phi(0)=0$, for three different 
times, $10, 30$ and $75$ ms and showing the 
averages over $z$. We plot $\rho(x,y;t)$ and 
contour plots. On the right panels we present contour 
plots of the averaged quantum phase, $\phi(x,y;t)$. 
The phase coherence of the condensates at each 
side of the barrier is clearly seen.}
\label{phase3dst}
\end{figure}

In Figs.~\ref{phase3d} and~\ref{phase3dst} 
we present full \tdgpetd simulations for a 
Josephson regime and a running phase mode self-trapped 
case, respectively. These two figures clearly 
show two relevant aspects of the problem. 
First, it is clear that during the full 
time evolution, which covers up to $t=80$ ms 
in the figure, the system remains mostly 
localized on the two minima of the potential. 
Therefore, the density has a two-peaked structure 
over the considered time period. Secondly, the atoms 
in each of the two wells remain to a large 
extent in a coherent phase during all times. 
This can be seen from the uniform color, constant 
phase, at each side of the barrier in the 
right panels of the figures. These two 
characteristics of the time evolution of 
the 3D Gross-Pitaevskii equation support the 
use of two-mode approximations. 

The modulation of the density profiles on 
the transverse direction is seen to be small, 
with a mostly constant quantum phase in 
the region populated by the atoms. This 
indicates that the transverse dynamics 
can be integrated out to a large extent, 
as is done in the 1D reductions discussed 
in Sec.~\ref{3D-1D-sec}. 

The Josephson dynamics, Fig.~\ref{phase3d}, 
is clearly seen in the small upper panels depicting 
$z(t)$ and $\delta\phi(t)$. They both oscillate 
with the same period but with a phase-shift of $\pi/2$. 

A self-trapped case is shown
in Fig.~\ref{phase3dst}. The atoms remain 
trapped mostly on the left side of the trap 
(they start with an imbalance of $z(0)=0.7$) 
and remain trapped in this potential-well 
during the considered time evolution. The 
coherence of the phase at each side of the 
potential barrier can also be appreciated 
in the figure, although here we should 
note that the right side of the barrier, 
being less populated, is concentrated 
on a smaller $(x,y)$ domain. 

\subsection{Comparison between the different models}

The \tdgpetd cases described above indicate 
that within the configuration considered here 
the two commonly employed two-mode models and 
1D equations are expected to be reasonable. 
In this section we present comparisons between 
the different approaches described in the 
previous sections: 1D reductions (\npse, \tdgpeod) 
and two-mode models, \sm and \im. 

\subsubsection{\tdgpetd vs 1D reductions: \tdgpeod and \npse}
\noindent

\begin{figure}[t]
\centering
\includegraphics[width=0.9\columnwidth,angle=0, clip=true]{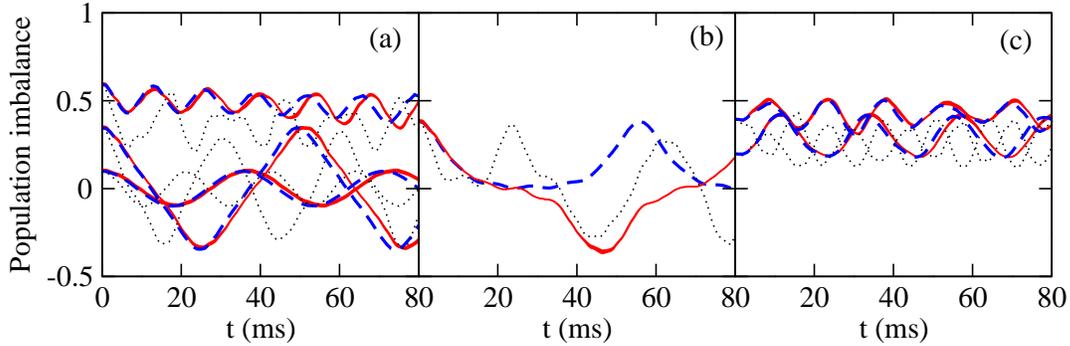} 
\caption[]{Dynamical evolution of the population 
imbalance, $z$, between both sides of the barrier 
for a single component condensate. Solid (red) line 
corresponds to the \tdgpetd, the dashed (blue) line 
to the \npse, and the dotted (black) stands for 
the \tdgpeod. Panel (a) contains $\delta\phi(0)=0$ 
cases, with $z(0)=0.1, 0.35$, and $0.6$. (b) 
Corresponds to the critical value, $z(0)=0.39$ 
and $\delta\phi(0)=0$. (c) Depicts two self-trapped 
cases with an initial $\delta\phi(0)=\pi$, 
with $z(0)=0.2$, and $0.4$. 
\label{comp}}
\end{figure}

In Fig.~\ref{comp} we present the time evolution of 
the population imbalance for the different dynamical 
conditions described in Sec.~\ref{sec:regimes}, 
i.e. Josephson, and self-trapping. We compare 
the full \tdgpetd (solid red) with the two 
previously described 1D reductions, \tdgpeod (dotted black) 
and \npse (dashed blue). 

First, we note that the dynamics emerging from 
the \tdgpetd is indeed similar to what 
was predicted by analyzing the \sm equations 
in Sec.~\ref{sec:regimes}. Qualitatively, 
the \tdgpetd simulations do follow the patterns 
predicted by the two-mode approximations. 
Lets us briefly describe each of the results:
 
\begin{itemize}
\item[a)] 
The first panel, (a), contains simulations performed 
with zero initial phase difference, i.e. Josephson 
oscillations and self-trapping cases. For the 
Josephson cases, $z(0)=0.1, 0.35$, the imbalance 
oscillates with a frequency which is mostly 
independent of the initial imbalance (for small 
imbalances). With $z(0)=0.1$ the oscillations 
are almost sinusoidal, while as we increase 
the initial imbalance their shape becomes 
more involved but remaining periodic. In the 
self-trapped case, $z(0)=0.6$, the atoms 
remain mostly on the initial side of the trap 
and there are short and small periodic oscillations as 
predicted by the two-mode models. At longer 
times, the imbalance is seen to decrease 
smoothly, implying a departure from the 
predicted two-mode dynamics~\cite{oursinprep}. 

\begin{figure}[t]
\centering
\includegraphics[width=0.85\columnwidth,
angle=0, clip=true]{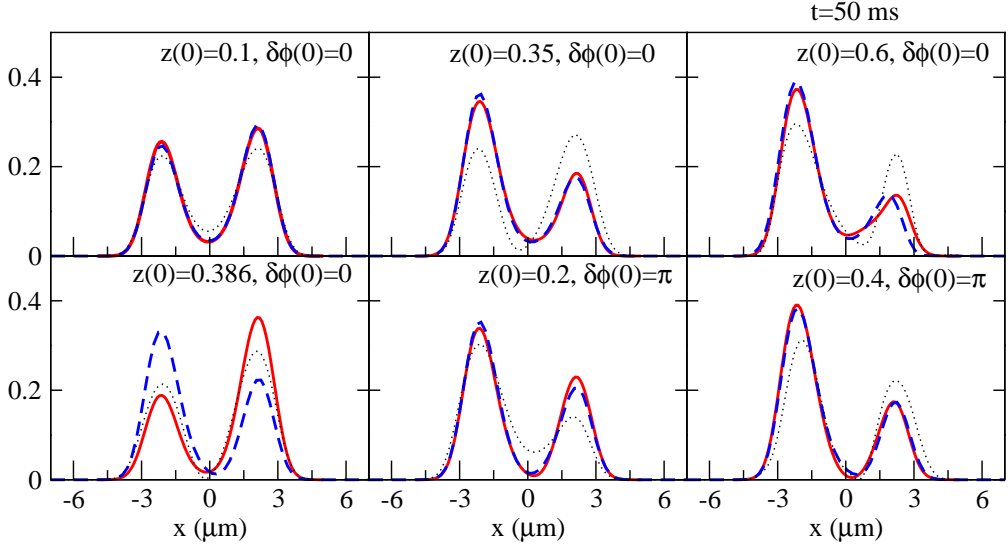} \caption[]{Snapshots 
of the axial density profiles, $\rho(x;t)$ $(\mu m)^{-1}$ at $t=50$ ms calculated 
by means of the \tdgpetd evolution (solid red line), the 
\npse (dashed blue line), and the \tdgpeod (dotted black 
line). The initial conditions correspond to the ones 
used to generate Fig.~\ref{comp}. } 
\label{profiles}
\end{figure}

The two 1D reductions give qualitatively similar 
results in most situations to \tdgpetd, but not 
quantitatively in all cases. The \npse is seen 
to reproduce very well the \tdgpetd in all the 
runs up to times near $\sim 40$ ms. Above those 
times, the period of oscillation predicted by 
the \npse is slightly shorter than the \tdgpetd one.  
%The \npse remains also mostly 
%two-mode on the entire time interval considered 
%providing a very accurate description of the 
%3D dynamics. 
The \tdgpeod on the contrary only 
captures the amplitude of oscillation in the 
Josephson cases, failing in all cases to give 
the same period as the \tdgpetd or the \npse. 
Moreover, the \tdgpeod departs notably from 
two-mode for the self-trapped case. It does 
predict self trapping, but more than two 
modes contribute to the time evolution. 

\item[b)]

Panel (b) is computed near the critical value of 
the full \tdgpetd, $z(0)=0.39$ for $\delta\phi(0)=0$. 
The \tdgpeod and \npse predict a critical initial 
imbalance close to the value predicted by the \tdgpetd.

\item[c)] 

Panel (c) contains two self trapped cases obtained 
with an initial $\delta\phi(0)=\pi$ and $z(0)=0.2$, 
and $0.4$. Notice that for $\delta\phi(0)=\pi$ 
the critical imbalance is smaller. The discussion is 
similar to the Josephson 
case, i.e. the \npse captures most of the dynamical features 
of the \tdgpetd while the \tdgpeod only provides a 
qualitative understanding of the problem. 
\end{itemize}
These results justify the use of the \npse in 
Ref.~\cite{Albiez05} to analyze their experiment. 

To further explore the quality of the 1D reductions, 
we present in Fig.~\ref{profiles} the density 
profiles in the $x$ direction after integrating 
the $y$ and $z$ components, 
$\rho(x;t)=\int_{-\infty}^\infty dy\,\int_{-\infty}^\infty\, dz\, |\Psi(x,y,z;t)|^2$ 
at $t=50$ ms. The agreement between the \npse and the 
\tdgpetd is very good in most situations, except for the 
critical case, as expected. In all cases the density 
profiles show a clear bi-modal structure. The \tdgpeod, as could 
be inferred from the previous results, does not 
predict the correct density profiles and, as seen 
in the self-trapped case, $(z(0)=0.6, \delta\phi(0)=0)$, 
do show the contribution of higher modes. The critical 
initial imbalance starting with no phase difference that we find 
numerically by means of the \tdgpetd is the same as 
found in Ref.~\cite{Ananikian2006}, $z_c=0.39$, and 
differs from the one reported in 
Ref.~\cite{Albiez05}, $z_c=0.5$. 

\begin{figure}[t]
\centering
\includegraphics[width=0.9\columnwidth,
angle=0, clip=true]{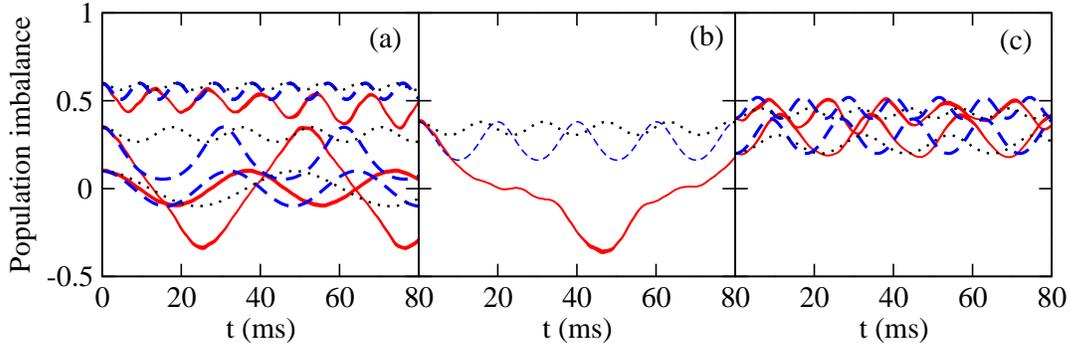} 
\caption[]{Dynamical evolution of the population 
imbalance between the two sides of the barrier 
for a single component condensate. The \tdgpetd 
(solid red) is compared to the \im (dashed blue) 
and the \sm (dotted black) results. The parameters 
entering in the two-mode descriptions are given in 
the text. Panel (a) contains runs for 
$\delta\phi(0)=0$, with $z(0)=0.1, 0.35$, and 
$0.6$. (b) Corresponds to the critical value 
for $z(0)=0.39$ and $\delta\phi(0)=0$. (c) 
Depicts two self-trapped states obtained by an 
initial $\delta\phi(0)=\pi$, with $z(0)=0.2$, and $0.4$. } 
\label{comp2}
\end{figure}

\subsubsection{\tdgpetd vs two-mode approximations, \sm and \im}
\noindent

As explained above, the use of 
two-mode models is suggested by the \tdgpetd 
results, see Figs.~\ref{phase3d} and~\ref{phase3dst}. 
What is, a priori, not clear, is whether the 
extra assumption used in deriving the \sm 
(which are the most commonly employed equations) 
will work for each specific double-well potential. 
As discussed in Sec.~\ref{sec:scaim}, the 
conditions of the Heidelberg experiment are such 
that the \sm predictions are not good. This does 
not mean that the dynamics is not two-mode but 
that the overlaps involving high powers of the 
two localized modes are not negligible as assumed 
in deriving the \sm equations. 

In Fig.~\ref{comp2} we compare \tdgpetd (solid red),  
the \sm (dotted black) and the \im (dashed blue) 
results using the parameters calculated microscopically 
from the ground and first excited state of the \tdgpetd. Both two-mode 
schemes predict the same phenomenology and thus 
qualitatively capture the dynamics of the system. 
At the quantitative level, however, the \im 
is clearly better. In the run with $z(0)=0.1$ 
and $\delta\phi(0)=0.$ (panel (a)), 
both the \sm and \im predict a similar behavior 
with the correct amplitude and oscillation period
close to the \tdgpetd one. As the 
imbalance is increased, e.g. (\cite{Albiez05} 
considers $z(0)=0.28$), the \sm fails to describe 
the correct period and predicts smaller 
amplitudes. This is analyzed in full detail 
in Ref.~\cite{Ananikian2006}.
The critical initial imbalances determined by both 
two-mode approaches are smaller than the \tdgpetd 
one. For the latter they predict a self-trapped 
case, see panel (b). Finally, for the self-trapped 
cases with $\delta\phi(0)=\pi$ (panel (c)) the 
\im give similar oscillation amplitudes 
with shorter periods than the \tdgpetd. The \sm 
fails both in reproducing the amplitudes and the 
periods.

%%%%%%%%%%%%%%%%%%%%%%%%%%%%%%%%%%%%%%%%%%
%% BINARY
%%%%%%%%%%%%%%%%%%%%%%%%%%%%%%%%%%%%%%%%%%

\section{Numerical solutions of the 3D Gross-Pitaevskii equations: 
binary mixture}

\label{sec:bmsp}

As discussed in Sec.~\ref{sec:system}, one 
feasible way of experimentally prepare binary 
mixtures of BECs is to consider a number of 
atoms populating the $m=\pm 1$ Zeeman components 
of an $^{87}$Rb $F=1$ spinor. The experimental 
observation of Josephson tunneling phenomena 
by the Heidelberg group seems 
to be possibly extended to trap both Zeeman 
components~\cite{private}. In this case the two components 
of the mixture have the same mass, 
$M \equiv m_a=m_b$, and equal intra-species 
interactions, $g_{aa}=g_{bb} \equiv g$. 
With respect to the inter-species interaction 
we will consider the case of $^{87}$Rb which 
implies $g_{ab}\sim g$.

The mean field \tdgpetd system of equations 
governing the dynamics of the three components 
of an $F=1$ spinor BEC can be written 
as~\cite{spinorBEC}, 
\begin{eqnarray}
i \hbar {\partial \psi_{\pm 1}\over \partial t} &=&
 [{\cal H}_s  + c_2(n_{\pm 1}+n_0- n_{\mp 1})]  \psi_{\pm 1}  
+ c_2  \psi_0^2 \psi^{*}_{\mp 1} \,, \nonumber \\ 
i \hbar  {\partial \psi_0\over \partial t} &=&
 [{\cal H}_s  + c_2(n_{1}+n_{-1})]  \psi_{0} 
+ c_2  2 \psi_{1} \psi_0^* \psi_{-1} \,, \label{dyneqs1}
\end{eqnarray}
with ${\cal H}_s=-\hbar^2/(2M)\, {\bm \nabla}^2 +V+c_0n$
being the spin-independent part of the Hamiltonian.
The density of the $m$-th component is given by
$n_{m}({\bf r})=|\psi_m({\bf r})|^2$, while 
$n({\bf r})=\sum_m|\psi_m({\bf  r})|^2$ is the total 
density normalized to the total number of atoms $N$.
The couplings are $c_0=4\pi\hbar^2(a_0+2a_2)/(3M)$ and
$c_2=4\pi\hbar^2(a_2-a_0)/(3M)$, where $a_0$ and $a_2$ 
are the scattering lengths describing binary elastic 
collisions in the channels of total spin 0 and 2, 
respectively. Their values for $^{87}$Rb are 
$a_0=101.8 a_B$ and $a_2=100.4 a_B$~\cite{vanKempen}. 
Since the spin-dependent coupling, $c_2$, is much 
smaller than the spin-independent one, $c_0$, and 
the total number of atoms that we will consider is 
relatively small $N=1150$, the population transfer 
between the different components can be 
neglected~\cite{Bruno2009}. Therefore, in our 
calculation the number of atoms in each sublevel 
remains constant in time allowing to treat the 
system as a real binary mixture of components 
$a$ and $b$. Comparing the system of 
Eqs.~(\ref{GP-mixture}) and (\ref{dyneqs1}) 
the value of the couplings can be read off, 
$g_{aa}=g_{bb}=c_0+c_2$ and $g_{ab}=g_{ba}=c_0-c_2$. 

Once the total number of atoms is fixed we want 
to investigate the Josephson-like dynamics 
for different number of atoms populating each 
component $N_a=f_a N$ and $N_b=f_b N$ and for 
different initial conditions $z_a(0)$, $z_b(0)$, 
$\delta\phi_a(0)$ and $\delta\phi_b(0)$. 

\begin{figure}[t]
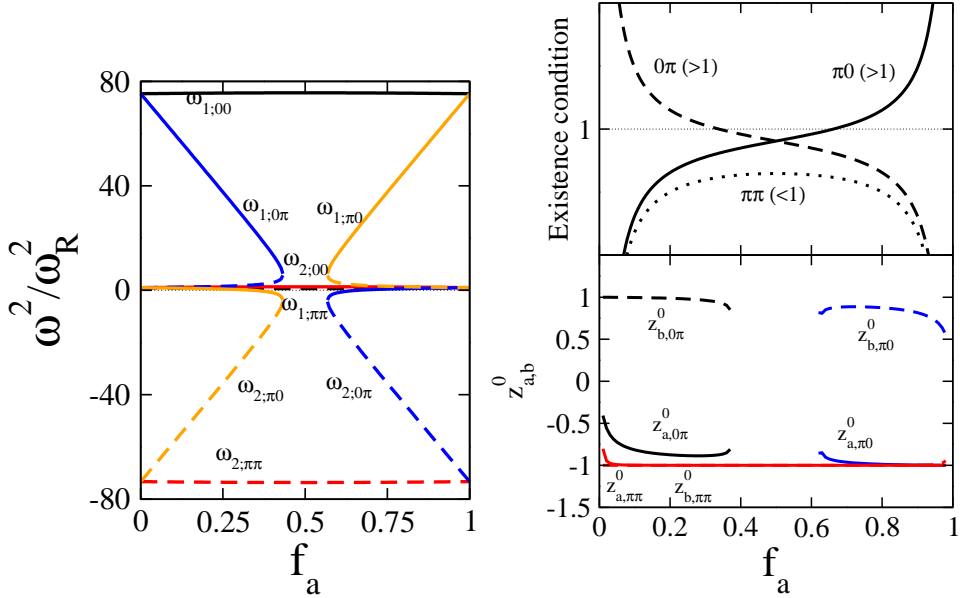

\centering
\includegraphics[width=.4\columnwidth,angle=0, clip=true]{fig11a.eps} 
\includegraphics[width=.4\columnwidth,angle=0, clip=true]{fig11b.eps} 
\caption[]{(left) Values of the frequencies, $\omega/\omega_R$, listed 
in Table~\ref{tabnm2} for the specific conditions considered 
in the numerical simulations as a function of the fraction 
of atoms in the $a$ component, $f_a$. The notation is as follows, 
$\omega_{i;\alpha\beta}$, with $i=1,2$ and $\alpha,\beta=0,\pi$. 
(right) The conditions for the existence of the non-trivial equilibrium 
points given in Eqs.~(\ref{eq:c1},~\ref{eq:c2},~\ref{eq:c3}), upper panel, 
as function of $f_a$ for the conditions described in the text. 
The lower panel contains the explicit equilibrium points $z_a^0, z_b^0$ 
as a function of $f_a$ obtained by solving equations~(\ref{eq:st2}). 
Note that each equilibrium point has a trivial partner 
which is obtained by flipping the sign of $z_a^0, z_b^0$. 
\label{fig:freqb}}
\end{figure}

The values of $\Lambda=NU/\hbar \omega_R$ and 
$\tilde{\Lambda}=N \tilde{U}/\hbar\omega_R$ 
are $\Lambda=74.278$ and $\tilde\Lambda=74.968$. 
With $\Lambda/\tilde\Lambda=0.99$. These are 
obtained from the microscopic 3D parameters computed 
in the scalar case, with the same total number of particles,
Sec.~\ref{sec:sc}. This is 
reasonable for the case we are considering where 
$g_{aa}=g_{bb}\sim g_{ab}$, which implies that the 
ground state wave functions for the GP equations of the mixture 
do not depend on $f_a$ and $f_b$ for a fixed total number 
of particles. This would certainly 
not be the case if $g_{aa}=g_{bb} \neq g_{ab}$, in such 
case one would need to recompute the ground state 
wave functions for $a$ and $b$ for each value of $f_a$. 

Following the discussion in Sec.~\ref{sec:drbm}, 
where the predictions of the \sm were discussed in 
detail, the system has the trivial equilibrium points, 
listed in Table~\ref{tabnm2} with $\beta=0.009$. In 
Fig.~\ref{fig:freqb} we show the values of the two 
eigenfrequencies for each of the trivial equilibrium 
points listed in Table~\ref{tabnm2} for the specific 
conditions described above. The figure shows a number 
of important features about the stability of the 
trivial equilibrium points. First, the 
$(z_a^0,\delta\phi_a^0,z_b^0,\delta\phi_b^0)=(0,0,0,0)$ 
is always stable regardless of the total polarization of 
the system (measured by $f_b-f_a$). Second, the 
$(z_a^0,\delta\phi_a^0,z_b^0,\delta\phi_b^0)=(0,\pi,0,\pi)$ 
mode is always unstable, as seen by the negative value 
taken by the square of the frequencies. Third, the 
$(z_a^0,\delta\phi_a^0,z_b^0,\delta\phi_b^0)=(0,0,0,\pi)$ 
mode should be stable for $f_a \lesssim 0.43$, 
correspondingly the 
$(z_a^0,\delta\phi_a^0,z_b^0,\delta\phi_b^0)=(0,\pi,0,0)$ 
is stable for  $f_b \lesssim 0.43$ and therefore there is a 
range of polarizations, given by $0.43 \gtrsim f_a \lesssim 0.57$ 
where the only trivial mode which is stable is the 
$(z_a^0,\delta\phi_a^0,z_b^0,\delta\phi_b^0)=(0,0,0,0)$. 

The non-trivial equilibrium points in this case can 
be obtained by analyzing the conditions given in 
Sec.~\ref{sec:drbm}. For $(\delta\phi_a^0,\delta\phi_b^0)=(0,0)$ 
there are no equilibrium points apart from the  
trivial one, due to $\Lambda\sim \tilde\Lambda$. 
In the other three cases there are non-trivial equilibrium 
points depending on the specific values of $f_a$. In 
Fig.~\ref{fig:freqb}(right) we analyze their existence. 
First, we note that there are non-trivial points 
corresponding to $(\delta\phi_a^0,\delta\phi_b^0)=(0,\pi)$ 
provided $f_a \lesssim 0.37$, correspondingly there are 
also equilibrium points for $(\delta\phi_a^0,\delta\phi_b^0)=(0,\pi)$ 
if $f_b\lesssim 0.37$. There is also a non-trivial 
equilibrium point for $(\delta\phi_a^0,\delta\phi_b^0)=(\pi,\pi)$ 
regardless of $f_a$. As can be seen in the figure, all these 
non-trivial equilibrium points correspond to fairly 
imbalanced conditions and can in most cases be understood 
in simple terms from the analysis of the scalar case. 
For instance, the equilibrium point for 
$(\delta\phi_a^0,\delta\phi_b^0)=(\pi,\pi)$ corresponds 
to $z_a^0\sim z_b^0\sim 1$ (or $-$1), which can be understood as 
having both components locked in a $\pi$-mode. Similarly, the 
equilibrium points in the $(0,\pi)$ or $(\pi,0)$ cases exist
whenever the most abundant component is populated enough to 
drive the dynamics close to being $\pi$ locked.

\subsection{\tdgpetd calculations: phase coherence and localization}

\begin{figure}[t]
\centering
\includegraphics[width=1\columnwidth,angle=0, clip=true]{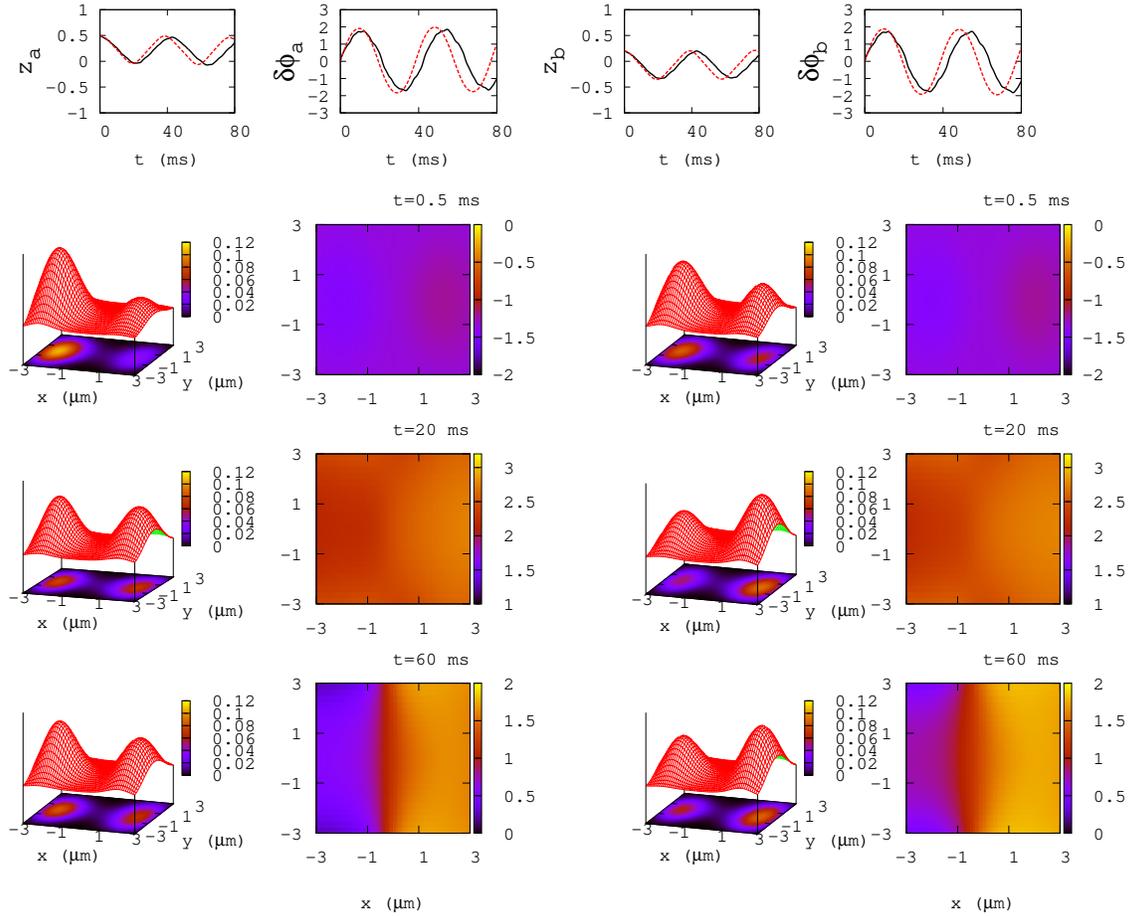} 
\caption[]{Full \tdgpetd calculations of the dynamics of a binary mixture 
with $z_a(0)=0.5$, $z_b(0)=0.2$, $\delta\phi_a(0)=0$, 
$\delta\phi_b(0)=0$, $f_a=0.25$ and $f_b=0.75$. The upper four 
plots depict, from left to right, $z_a(t)$, $\delta\phi_a(t)$,
 $z_b(t)$, and $\delta\phi_b(t)$ in solid black compared to 
the \im prediction, dashed-red . Then each row contains from 
left to right: 3D depictions complemented 
by contour plots of $\rho_a(x,y;t)$, a contour plot of the 
averaged phase $\phi_a(x,y;t)$, 3D depictions complemented 
by contour plots of $\rho_b(x,y;t)$, and a contour plot of 
the averaged phase $\phi_b(x,y;t)$. Each row corresponds 
to a different time, .5 ms (upper), 20 ms (middle) and 
60 ms (lower), respectively. 
\label{3dbin4}}
\end{figure}

The numerical solutions of the \tdgpetd presented 
in Sec.~\ref{sec:sc} for the single component case 
showed two features. First, the atoms remained 
mostly localized in the two minima of the potential 
well and secondly, each group of atoms 
had to a large extent the same quantum phase. This, 
clearly supported the picture of having two BEC, 
one at each side of the barrier, with a well 
defined phase at each side during the dynamical 
evolution. Essentially those are the premises 
used to derive the two-mode models, both for 
single component and for binary mixtures, as 
we did in Sec.~\ref{sec:tw}. 

As in the scalar case, our exact \tdgpetd numerical 
solutions of the dynamics of the binary mixture in 
several initial conditions of population imbalances and 
phase differences show two distinctive features, see 
Fig.~\ref{3dbin4}. First, the density of atoms for 
each component is always bi-modal, with the two 
atom bunches centered around the minima of the 
potential well. Secondly, the phase of the wave 
function is mostly constant for each species at 
each side of the potential trap. Thus, we find 
that the \tdgpetd does predict the dynamics to 
be mostly bi-modal also for the binary mixture case. 

At the end of the section we will consider some 
deviations from the bi-modal behavior that are found 
in very specific conditions, e.g. for very large 
population imbalances and also when analyzing a case 
with $g_{ab}\neq g_{aa}=g_{bb}$.

\subsection{Small oscillations around $z_{a,b}^0$ and $\delta\phi_{a,b}^0=0$}

\begin{figure}[t]
\centering
\includegraphics[width=0.85\columnwidth,angle=0, clip=true]{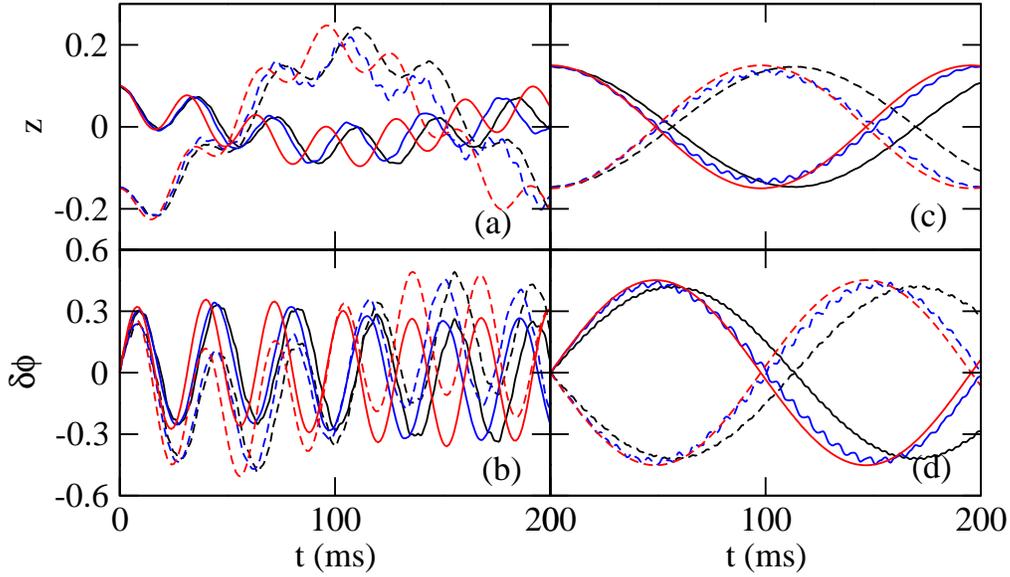} 
\caption[]{Behavior of the population imbalance, $z_a(t)$ (solid lines), 
and $z_b(t)$ dashed lines, and phase difference, 
$\delta\phi_a(t)$ (solid lines) and $\delta\phi_b(t)$ (dashed lines), 
computed using \tdgpetd (black lines),  
\npse (blue lines), and \im (red lines) in a polarized 
case, $f_a=0.8$, left, and a zero polarization case, 
$f_a=0.5$, right, respectively. The initial conditions 
are $z_a(0)=0.1$, $z_b(0)=-0.15$ 
and $\delta\phi_a(0)=\delta\phi_b(0)=0$ for the left panels, and 
$z_a(0)=-z_b(0)=0.15$ and $\delta\phi_a(0)=\delta\phi_b(0)=0$ for 
the right panels. 
\label{fig2new}}
\end{figure}

\begin{figure}[t]
\centering
\includegraphics[width=0.85\columnwidth,angle=0, clip=true]{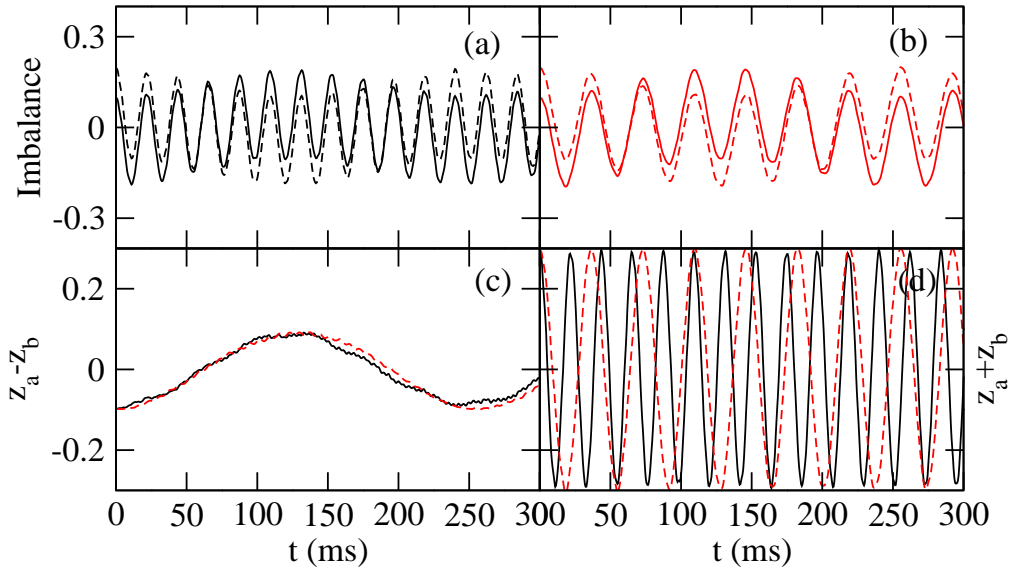} 
\caption[]{Behavior of the population imbalance in \npse (red) 
and \gpod (black) simulations in the zero magnetization case, $f_a=f_b$. The initial 
conditions are $z_a(0)=0.1$, $z_b(0)=0.2$ and $\delta\phi(0)=0$. 
The upper panels correspond to (a) $z_a(t)$ (solid line) 
and $z_b(t)$ (dashed line) obtained 
with the \tdgpeod equations, (b) $z_a(t)$ and $z_b(t)$ obtained 
with the \npse equations, (c) behavior of $z_a(t)-z_b(t)$ for 
\tdgpeod (solid) and \npse (dashed), and (d) behavior of  $z_a(t)+z_b(t)$. 
\label{fig1new}}
\end{figure}

\begin{figure}[t]
\centering
\includegraphics[width=0.8\columnwidth,angle=0, clip=true]{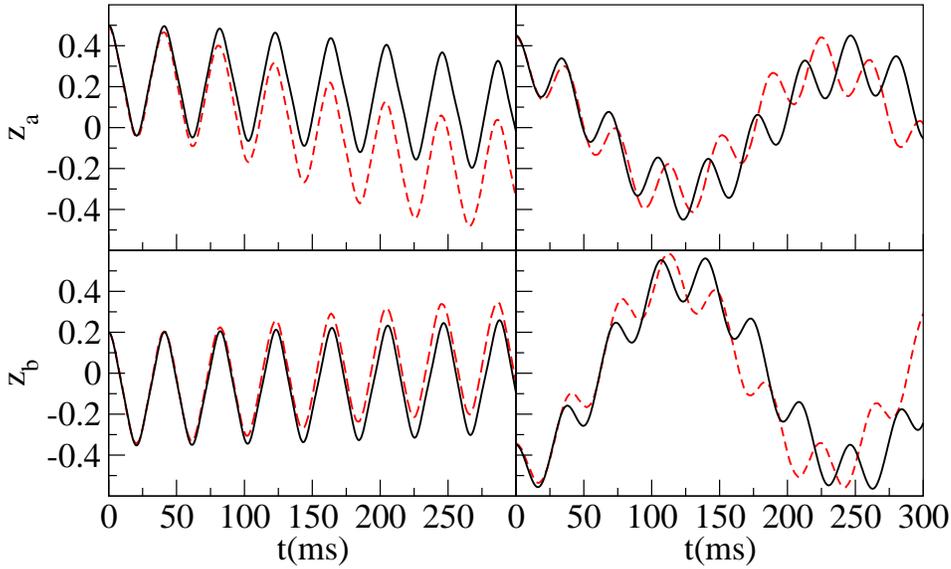} 
\caption[]{(left panels) Evolution of the population imbalance 
of each component for a binary mixture with 
$f_a=0.25$. The upper panel shows $z_a(t)$, and the 
bottom panel $z_b(t)$. The solid (black) line 
corresponds to the \im model and the dashed (red) 
line to the \npse. The initial conditions are 
$z_a(0)=0.5$, $z_b(0)=0.2$, $\delta \phi_a(0)=\delta \phi_b(0)=0$.
(right panels) As in the left panel, but with $f_a=0.6$ and 
initial conditions $z_a(0)=0.45$, $z_b(0)=-0.35$, 
$\delta \phi_a(0)=\delta \phi_b(0)=0$. }
\label{fig-24joseph2}
\end{figure}

The two predictions of the \sm described in 
Sec.~\ref{sec:drbm} are confirmed by the 
\npse and \gpod simulations as can be seen 
in Figs.~\ref{fig2new} and~\ref{fig1new}. 
In Fig.~\ref{fig2new} (left panels) we 
consider a very polarized case, $f_a=0.8$. 
As expected from the two-mode analysis 
the dynamics of the most populated 
component should to a large extent decouple from 
the less populated one and perform fast Josephson 
oscillations with a frequency close to the 
corresponding one for the scalar case, 
$\omega_J=\omega_R\sqrt{1+\Lambda}$. The \tdgpetd 
simulation is seen to confirm the above and follow 
closely the predictions of the \im. The less abundant 
component is strongly driven by the most populated one 
and shows an anti-Josephson behavior as described in 
Ref.~\cite{Bruno2009}.  

Another prediction is related to the behavior 
of $z_a+z_b$ and $z_a-z_b$ in the non-polarized 
case, $f_a=f_b$. As explained in Sec.~\ref{sec:drbm}, 
in this case the difference, $z_a-z_b$, should enhance 
the long mode which oscillates with the Rabi frequency 
of the system, while the sum $z_a+z_b$ should mostly 
oscillate with the Josephson frequency. In the right 
part of Fig.~\ref{fig2new} we present the extreme 
case when $z_a(0)=-z_b(0)$ computed with \tdgpetd, 
\npse and \im. In this case, both population imbalances 
and phase differences oscillate mostly with the Rabi 
frequency of the system, keeping during the time 
evolution $z_a+z_b\sim 0$. 

As seen in Fig.~\ref{fig1new} both 1D 
reductions produce qualitatively similar physics. 
The only important difference is that 
the frequency of the Josephson oscillations is 
higher in the \gpod, as occurred already for 
the single component, see Sec.~\ref{sec:sc}. 

Interestingly, they predict different 
Josephson oscillations while the Rabi frequencies are similar. 
In panel (c) of Fig~\ref{fig1new} the long oscillation corresponding 
to the Rabi mode is seen to agree well with the 
corresponding long oscillation seen in the right 
panels of Fig.~\ref{fig2new}. The Josephson-like 
oscillations of binary 
mixtures of spinor $F=1$ $^{87}$Rb BECs around the 
$(z_a^0,\delta\phi_a^0,z_b^0,\delta\phi_b^0)=(0,0,0,0)$ 
are therefore essentially controlled by two 
frequencies, $\omega_R$ and $\omega_J$. 

As a general statement, in the conditions of 
the Heidelberg experiment, as 
occurred for the scalar case, the \im produces 
more reliable results than the \sm model, which are not 
shown in the figures. Notice that the parameters that 
we use for the \im are extracted from the \tdgpetd 
calculation as given in Sec.~\ref{sec:sc}. Other 
representative cases with 
$(\delta\phi_a(0),\delta\phi_b(0))=(0,0)$ but with 
larger initial imbalances, $z_i(0)\sim 0.5$ are shown 
in Fig.~\ref{fig-24joseph2}. On the left side of the 
figure we show the population imbalance of each component 
for a simulation with $f_a=0.25$. In this case the dynamics 
is controlled by $\omega_J$. The panel on the right depicts 
a simulation with $f_a=0.6$ and close to opposite initial 
population imbalances. In this case, both frequencies 
$\omega_J$ and $\omega_R$ show up in the evolution. The \im 
provides a satisfactory description of the dynamics.

\subsection{Small oscillations around $z^0_{a,b}$, 
$\delta\phi^0_{a}=0$ and $\delta\phi^0_{b}=\pi$}

\begin{figure}[t]
\centering
\includegraphics[width=0.85\columnwidth,angle=0, clip=true]{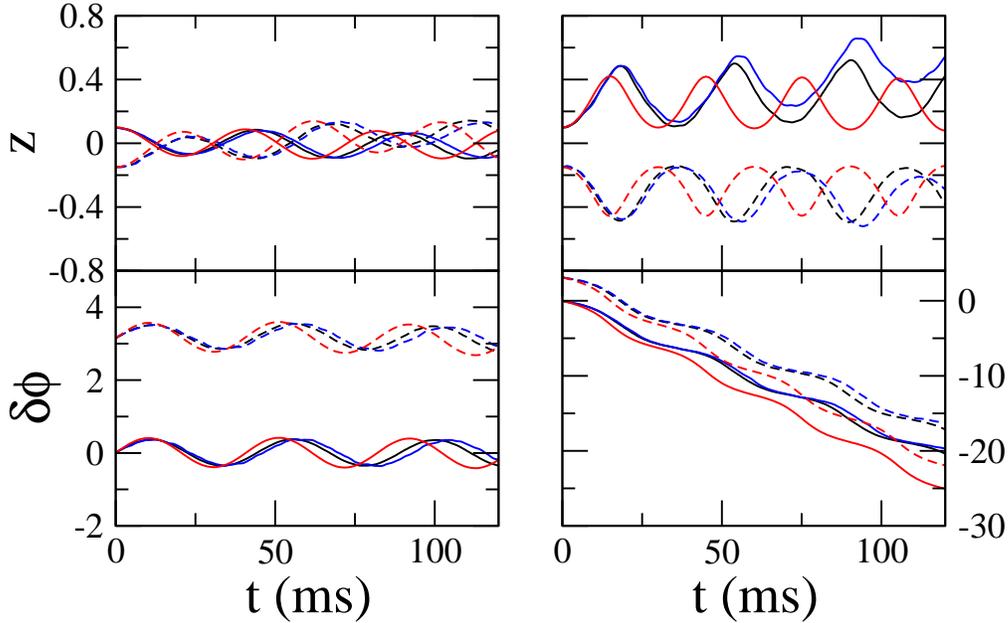} 
\caption[]{Two simulations with the same initial conditions, $z_{a}(0)=0.1$,
  $z_b(0)=-0.15$, $\delta\phi_{a}(0)=0$ and $\delta\phi_{b}(0)=\pi$ but 
with different compositions of the mixture. The case
on the left has $f_a=0.2$ while the case on the right $f_a=0.8$. The 
blue lines are obtained by means of a full \tdgpetd, the 
black lines are the \npse results, and the red lines are the 
\im results. Solid and dashed lines correspond to 
the $a$ and $b$ components, respectively.}
\label{figbin0p1}
\end{figure}

\begin{figure}[t]
\centering
\includegraphics[width=0.85\columnwidth,angle=0, clip=true]{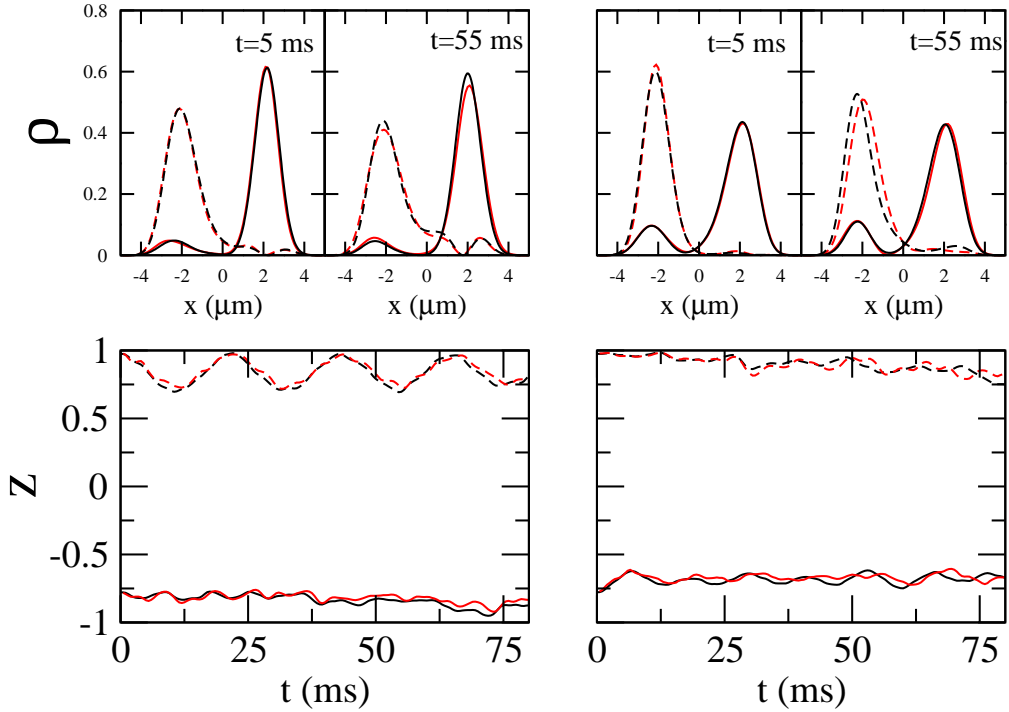} 
\caption[]{Two simulations with the same initial conditions, $z_{a}(0)=-0.78$,
  $z_b(0)=0.99$, $\delta\phi_{a}=0$ and $\delta\phi_{b}=\pi$, but 
with different composition. The case
on the left has $f_a=0.1$ while the case on the right $f_a=0.9$. The 
red lines are obtained by means of a full \tdgpetd while the 
black lines are the \npse results. Solid and dashed lines correspond to 
the $a$ and $b$ components, respectively.}
\label{figbin0p2}
\end{figure}

As explained above, for these conditions there 
can exist up to three stationary points depending 
on the specific value of $f_a$ considered. The 
trivial equilibrium point exists provided 
$f_a\lesssim 0.43$, see Fig.~\ref{fig:freqb}. This 
prediction of the two-mode models is observed both 
in the \tdgpetd and \npse as it can be seen in 
Fig.~\ref{figbin0p1}. In the figure, we consider a 
simulation with $z_a(0)=0.1$, $z_b(0)=-0.15$, and 
$f_a=0.2<0.43$ (left panels). The population imbalance (upper panel) of both
components oscillates in the usual Josephson regime. At the same time, the
phase difference oscillates with its characteristic phase-shisft of $\pi/2$
with respect to the imbalance (lower panel). The phase of the 
$a$ component oscillates around $\delta\phi_a=0$ 
while $\delta\phi_b$ does oscillate around 
$\delta\phi_b=\pi$. 

A completely different picture emerges when the 
fraction of atoms in both components is exchanged, $f_a=0.8>0.43$ (right
panels), with most of the atoms populating 
the $a$ component. In this case, the oscillation 
amplitude is large, both components remain 
trapped on their original sides and the phase 
difference becomes unbounded. This should be 
considered as a genuine effect of the binary 
mixture as each component follows a running 
phase mode at each side of the potential barrier. 

The comparison between the \npse and \tdgpetd 
is very satisfactory. The \npse captures 
almost completely the dynamics up to times 
of 100 ms. In all cases, the \npse reproduces 
correctly both the phase difference and population 
imbalance. The only sizeable 
discrepancies occur for times $\gtrsim 80$ ms 
in the run without equilibrium point (right panel).

The \im gives a good qualitative picture of both 
cases but fails to provide predictions as accurate 
as the \npse, as happened in the scalar case, see 
for instance Figs.~\ref{comp} and ~\ref{comp2}. In 
particular the predicted periods of oscillation are 
much longer than the actual ones.

An example of simulations around non-trivial 
equilibrium points is presented in Fig.~\ref{figbin0p2}.
As explained previously, these involve very large 
and opposite initial population imbalances  for both components. 
In Fig.~\ref{figbin0p2} we consider 
a case with initial conditions very close to the predicted equilibrium 
point using the standard two-mode, and described 
in Fig.\ref{fig:freqb}, $z_a(0)=-0.78$, and 
$z_b(0)=0.99$, with $f_a=0.1$. Also in the same 
figure we consider a similar run but with $f_a=0.9$. 
In both cases the \npse and \tdgpetd predict a very 
similar dynamics. These simulations will be discussed 
again in Sec.~\ref{btm} as they exhibit effects which 
clearly go beyond a two-mode approximation.

\subsection{Small oscillations around $z^0_{a,b}$ and $\delta\phi^0_{a,b}=\pi$}

\begin{figure}[t]
\centering
\includegraphics[width=0.85\columnwidth,angle=0, clip=true]{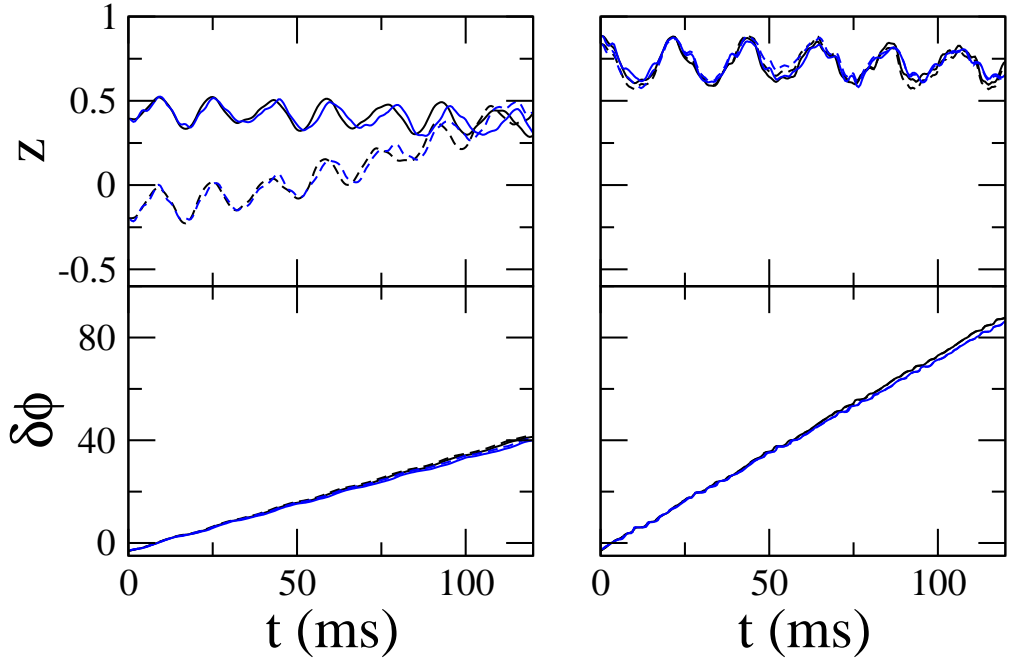} 
\caption[]{Two simulations corresponding to (left) $z_{a}(0)=0.4$,
  $z_b(0)=-0.2$, $\delta\phi_{a}(0)=\pi$, $\delta\phi_{b}(0)=\pi$ 
and $f_a=0.9$, and (right) $z_{a}(0)=0.9$,
  $z_b(0)=0.85$, $\delta\phi_{a}(0)=\pi$, $\delta\phi_{b}(0)=\pi$ and $f_a=0.9$
The blue lines are obtained by means of a full \tdgpetd while the 
black lines are the \npse results. Solid and dashed lines correspond to 
the $a$ and $b$ components, respectively.}
\label{figbinpp1}
\end{figure}

The trivial equilibrium point is not stable in 
the considered conditions as seen in 
Fig.~\ref{fig:freqb}. The non-trivial one, however, 
is only attainable if extremely imbalanced configurations 
for both components are considered. This case would 
correspond essentially to having both components in 
a $\pi$ mode state, which in our conditions only 
exists for $z~\sim 1$ as can be seen in the blue spots in 
panel (c) of Fig.~\ref{fig:tmscalar-r}. In Fig.~\ref{figbinpp1}
we present two simulations with different initial conditions. 
First, we consider a simulation with $z_a(0)=0.4$ and 
$z_b(0)=-0.2$, with $f_a=0.9$. The behavior is understood in simple 
terms, the most populated component remains self-trapped while 
the other component is forced by the other one. The phase 
evolves unbounded. The figure again contains \tdgpetd and 
\npse simulations.

The second simulation (right panels) is closer to a non-trivial equilibrium 
point, we consider $z_a(0)=0.9$ and $z_b(0)=0.85$ with $f_a=0.9$. 
In this case, both components remain self trapped, the 
phase difference is unbounded, but we do not get the expected 
behavior of two $\pi$ modes because the 
initial imbalances are not close enough to $z^0\sim 1$. 

\subsection{Effects beyond two-mode}
\label{btm}

Most of the dynamics described in the previous 
sections can to a large extent be understood 
within the two-mode models developed in 
Sec.~\ref{sec:tw}. There are, however, a 
number of situations where the two-mode fails. 
Some are a direct consequence of having two 
components evolving in the same double-well 
potential, others are due to having initial 
configurations, mostly with large initial 
imbalances, producing situations 
where the atom-atom interaction energy per atom 
is comparable to the gap between the first 
excited state and the second/third excited 
states. 

We can distinguish two different cases: (a) 
involving excitations along the coordinate 
which contains the barrier, (b) involving 
excitations of the transversal coordinates. 

An example of (a) is seen in Fig.~\ref{figbin0p2}. 
There, as clearly seen in the density profiles along 
the $x$ direction, the two-mode approximation is 
clearly not valid. The simplest way of seeing this 
is by noting the zero in the density of one of the components 
at $x\sim 2 \mu$ m. This effect beyond two-mode is 
well taken care of by the \npse which reproduces the 
density profile quite well during most of the time 
evolution considered in the simulation. Thus, the 
excitations of higher modes along the direction which 
has not been integrated out in the 1D reduction do 
not pose a great difficulty to the 1D reductions. 

\begin{figure}
\centering
\includegraphics[width=1\columnwidth,
angle=0, clip=true]{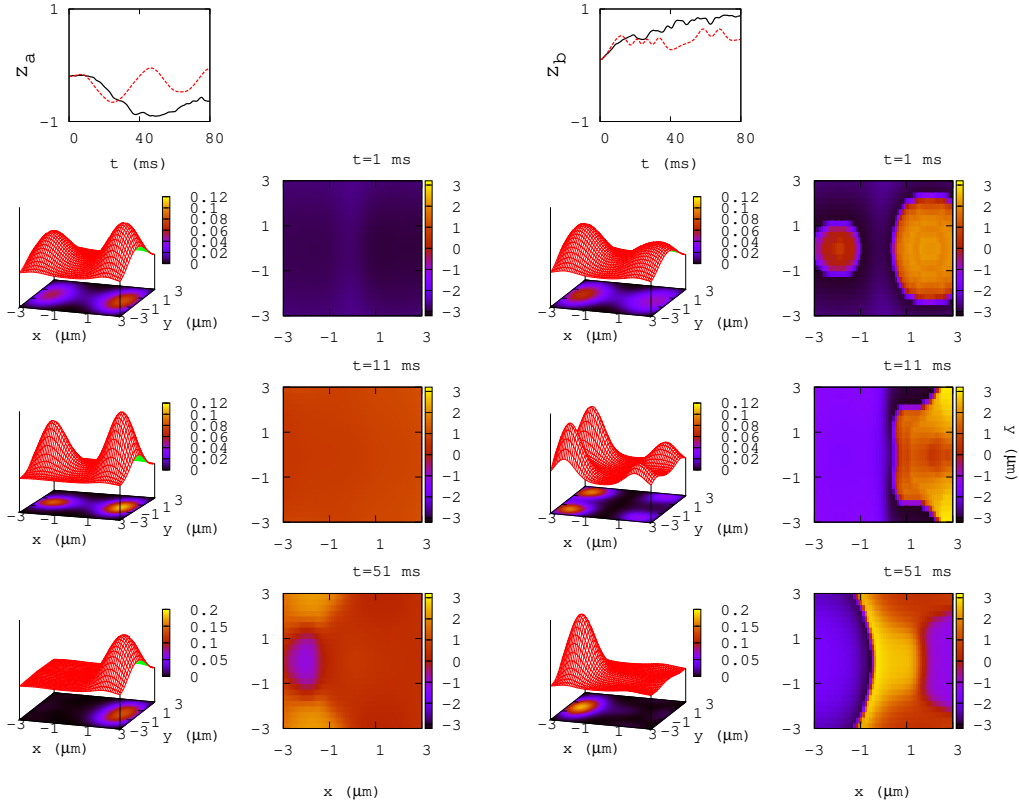}
\caption[]{Full \tdgpetd calculations of the dynamics 
of a binary mixture with $z_a(0)=-0.2$, $z_b(0)=0.1$, 
$\delta\phi_a(0)=0$, $\delta\phi_b(0)=0$, $f_a=0.7$ and 
$f_b=0.3$. As explained in the text in this case, 
$g_{aa}=g_{bb}$ and $g_{ab}=2.3 g_{aa}$. The upper 
two plots depict $z_a(t)$ (left) 
and $z_b(t)$ (right). Then each row contains from 
left to right: 3D depictions complemented by contour 
plots of $\rho_a(x,y;t)$, a contour plot of the averaged 
phase $\phi_a(x,y;t)$, 3D depictions complemented 
by contour plots of $\rho_b(x,y;t)$, and a contour plot of 
the averaged phase $\phi_b(x,y;t)$. Each row corresponds 
to a different time, 1 ms (upper), 11 ms (middle) and 
51 ms (lower), respectively. In all cases, solid black 
lines are computed with \tdgpetd and dashed red ones with 
\im. }
\label{fig:btw1}
\end{figure}

The second type, (b), of effects beyond two-mode 
involve excitations of the transverse components. These 
effects are present in any binary mixture calculation 
whenever the intra- and inter-species interactions 
are not equal.  To enhance 
this effect, and also to explore the 
interesting symmetry breaking phenomena described in 
Ref.~\cite{Satija2009}, we consider a case with $g_{aa}=g_{bb}$, 
but with $g_{ab}=g_{ba}=2.3 g_{aa}$. Therefore, now the 
inter-species interaction strength is larger than 
the intra-species one. The two-mode prediction for this 
case, \sm, which was analyzed in Ref.~\cite{Satija2009} shows 
a large symmetry breaking pattern during the time 
evolution of the system. In Fig.~\ref{fig:btw1} we 
consider a full \tdgpetd simulation of a representative example
with $z_a(0)=-0.2$, 
$z_b(0)=0.1$, $\delta\phi_a(0)=\delta\phi_b(0)=0$, 
and $f_a=0.7$.

The qualitative prediction of the \im also shows the symmetry breaking, and 
the two components do separate from each other and mostly 
concentrate on one of the wells as time evolves. But, as 
it can be seen in the 3D depictions of $\rho(x,y;t)$ at 
three different times, the evolution of the system departs 
almost from the beginning from the two-mode. At $t=1$ ms 
we have the density distributions of each component corresponding 
to a small initial imbalance. Then at $t=11$ ms, we can already 
see that the most populated component is expelling the other 
one from the minima of the potential. This fact can be appreciated as a four
peaked  
distribution, $\rho_b(x,y;t)$. After that, each of the 
components start to accumulate on their original sides following 
qualitatively the prediction of the \im and thus presenting 
the symmetry breaking pattern discussed in 
Ref.~\cite{Satija2009}. The two-mode approximation is in 
this case broken for a short period of time, when the 
first modes along the transverse directions are excited 
due to the large inter-species interaction. 

\section{Conclusions}
\label{sec:sum}

We have presented a thorough investigation of the mean-field 
dynamics of a binary mixture of Bose-Einstein 
condensates trapped in a double-well potential. 
The rich dynamical regimes which take place in binary 
mixtures, like double self-trapped modes, Josephson 
oscillations, or zero and $\pi$ bound phase modes, have been 
scrutinized by performing full \tdgpetd simulations covering 
all the relevant initial conditions. The 3D numerical 
solutions of the Gross-Pitaevskii equations have been used as a benchmark to
critically discuss the validity of the most common 
1D reductions of the GP equations, \tdgpeod and \npse, and 
the often employed simple two-mode reductions, \sm and \im. 

The full 3D solutions of the binary mixture have 
shown to have a large amount of phase coherence 
and localization at each side of the potential barrier 
for both components, predicting a dynamics which is mostly bi-modal. This feature permits to 
speak of Bose-Einstein condensates at each side of 
the barrier, where the atoms mostly share a common 
phase, and to support the use of two-mode approximations, which analytical
solutions allow to gain physical insight into the problem.

To fix the conditions of the dynamics, we have focused 
in one particular setup that corresponds to a natural 
extension of the experiments reported in Ref.~\cite{Albiez05}: 
the case of a binary mixture made by populating two 
of the Zeeman states of an $F=1$ $^{87}$Rb condensate.
As discussed in the present paper, this setup already allows 
to observe and characterize a large variety of phenomena 
which are genuine of the binary mixture, e.g. anti-Josephson 
oscillations in highly polarized cases, long Rabi-like 
oscillating modes, zero and $\pi$ locked modes, etc. 

For the sake of completeness and to better frame 
the physics of the binary mixture we have provided 
a detailed description of the single component 
dynamics, with explicit expressions for 
all the commonly employed approximations to the 
3D mean field Gross-Pitaveskii equation. The natural extension of the 
latter to the binary mixture, i.e. \sm and \im 
equations and 1D reductions, have been 
consistently derived providing a self-contained reference, easy to read, with
all the relevant formulae used in the article.

The standard two-mode model, with its 
microscopic parameters computed with the 
\tdgpetd, has been used to reexamine the existence and stability of 
the different regimes that can occur in 
both single component and binary  mixture 
condensates, describing the Josephson 
oscillations and the macroscopic quantum 
self-trapping, including running phase modes 
and zero- and $\pi-$modes. 

The comparisons between the two-mode models and 
the numerical solutions of the \tdgpetd show 
an excellent agreement for conditions close 
to the stable stationary regimes predicted by the 
two-mode models. As we depart from those stable 
points, the \sm fails to provide a quantitative agreement 
with the results obtained with the \tdgpetd equations. 
The range of validity of the \im is much larger, 
fully capturing the dynamics of single and binary mixtures 
for a larger set of initial conditions. 

The two most commonly employed dimensional 
reductions of the \tdgpetd, the \gpod and \npse, have been 
shown to differ substantially among each other, with the 
\npse being clearly in much better agreement with the 
original 3D dynamics in a broader set of conditions. 
In general, the \gpod describes essentially the 
correct physics but quantitatively far from the \tdgpetd predictions. Also, 
for self-trapped cases already in the single component 
case, it departs from the two-mode behavior earlier than 
the \tdgpetd or the \npse. The agreement between the \npse 
and the full 3D dynamics is astonishingly good both for single 
component and the considered binary mixtures, where 
the intra- and inter-species are very similar and the \npse 
equations are particularly easy to handle. This agreement 
is not only seen on fully integrated magnitudes, for instance population 
imbalances, but also on the density profiles predicted 
along the direction hosting the barrier. 

We have also considered two situations where the two-mode approximation 
fails. This is naturally due to the excitation of higher modes. 
Two different cases have been described, first the excitation of 
modes in the direction of the barrier and secondly, excitation 
of modes in the transverse direction. The \npse has been shown to 
capture perfectly the excitations along the barrier direction, 
reproducing the integrated density profiles obtained with the 
\tdgpetd. The second case has been studied in a simulation performed 
with different intra- and inter-species, which can be achieved in principle 
experimentally through Feshbach resonance modulation of the 
scattering lengths. In this case, the dynamics of the less populated 
component in each side of the trap departs notably from the 
two-mode with clear excitations of transverse modes, seen already 
in the density profiles along a transverse direction. 

The present article is intended both to motivate the experimental 
effort to study binary mixtures of BECs, where we have shown that 
a large variety of phenomena related to phase coherence and 
localization can be observed, and to serve as a tool in the analysis 
of such experiments providing a concise and self-contained  
derivation of the most commonly used models.

\section*{Acknowledgments}
We thank J. Martorell and M. Oberthaler for useful 
discussions. B.J-D. is supported by 
a CPAN CSD 2007-0042 
contract, Consolider Ingenio 2010.
This work is also supported by the
Grants No. FIS2008-00421, FIS2008-00784, 
FIS2008-01236 and 2005SGR-00343,  SGR 2009-0985
from Generalitat de Catalunya and 
Consolider Ingenio 2010 QOIT.

\end{document}